\documentclass[11pt,letterpaper]{article}
\usepackage[T1]{fontenc}
\usepackage[utf8]{inputenc}
\usepackage{lmodern}
\usepackage[margin=1in]{geometry}
\usepackage{amsmath,amssymb,amsthm,mathtools}
\usepackage{microtype,booktabs,array,placeins,aliascnt,needspace}
\usepackage[ruled,vlined,linesnumbered]{algorithm2e}
\usepackage{enumitem}
\usepackage{tikz}
\usetikzlibrary{arrows.meta,patterns}
\usepackage[colorlinks=true,linkcolor=blue,citecolor=blue,urlcolor=blue]{hyperref}
\usepackage[nameinlink,noabbrev,capitalise]{cleveref}
\hypersetup{pdftitle={l-infinity norm SVP in O-tilde(3 to the n)},pdfauthor={Yiming Gao}}
\setlist[enumerate]{itemsep=3pt,topsep=5pt}
\SetAlFnt{\small}
\SetAlCapFnt{\small\bfseries}
\SetAlCapNameFnt{\small\bfseries}
\DontPrintSemicolon
\newtheorem{theorem}{Theorem}[section]
\newaliascnt{lemma}{theorem}
\newtheorem{lemma}[lemma]{Lemma}
\aliascntresetthe{lemma}
\newaliascnt{proposition}{theorem}
\newtheorem{proposition}[proposition]{Proposition}
\aliascntresetthe{proposition}
\newaliascnt{corollary}{theorem}
\newtheorem{corollary}[corollary]{Corollary}
\aliascntresetthe{corollary}
\theoremstyle{definition}
\newaliascnt{definition}{theorem}

\aliascntresetthe{definition}
\theoremstyle{remark}
\newaliascnt{remark}{theorem}

\aliascntresetthe{remark}
\crefname{theorem}{Theorem}{Theorems}
\crefname{lemma}{Lemma}{Lemmas}
\crefname{proposition}{Proposition}{Propositions}
\crefname{corollary}{Corollary}{Corollaries}
\crefname{section}{Section}{Sections}
\crefname{appendix}{Appendix}{Appendices}
\crefname{figure}{Figure}{Figures}
\DeclareMathOperator{\vol}{vol}

\DeclareMathOperator{\sinc}{sinc}
\DeclareMathOperator{\Var}{Var}
\DeclareMathOperator{\lab}{lab}
\DeclareMathOperator{\poly}{poly}
\newcommand{\R}{\mathbb R}
\newcommand{\Z}{\mathbb Z}

\newcommand{\F}{\mathbb F}
\newcommand{\E}{\mathbb E}
\newcommand{\Prb}{\mathbb P}
\newcommand{\ind}{\mathbf 1}

\newcommand{\infnorm}[1]{\lVert #1\rVert_\infty}
\newcommand{\eunorm}[1]{\lVert #1\rVert_2}

\newcommand{\OO}{\widetilde O}
\newcommand{\G}{\mathcal G}
\newcommand{\List}{\mathcal X}
\newcommand{\Fail}{\mathsf{fail}}
\newcommand{\Build}{\mathsf{Build}}

\newcommand{\Best}{v_{\mathrm{best}}}
\newcommand{\Ared}{A_{\mathrm{red}}}
\newcommand{\Lcap}{L_{\max}}
\newcommand{\Wcap}{W_{\max}}
\newcommand{\bits}{\operatorname{size}}
\newcommand{\LeanRepoURL}{\url{https://github.com/Gym0000/linfty-svp-lean}}

\definecolor{cIn}{RGB}{105,105,105}
\definecolor{cMid}{RGB}{28,84,168}
\definecolor{cDiff}{RGB}{214,96,0}
\definecolor{cCov}{RGB}{20,135,115}
\newcommand{\FigGamma}{0.75}
\tikzset{
  paperfig/.style={line cap=round,line join=round,
    >={Stealth[length=3.6pt,width=2.9pt]}},
  lab/.style={font=\scriptsize,inner sep=1.5pt},
  axis/.style={->,cIn!85!black,line width=0.4pt},
  tick/.style={cIn!85!black,line width=0.4pt}}

\newcommand{\PlaneBox}{\useasboundingbox (-1.36,-1.34) rectangle (1.46,1.46);}
\newcommand{\PlaneSquare}{\fill[cIn!7] (-1,-1) rectangle (1,1);}
\newcommand{\PlaneStrip}[2]{%
  \begin{scope}
  \clip (-1,-1) rectangle (1,1);
  \fill[#1] (-2,{-2-\FigGamma}) -- (2,{2-\FigGamma})
    -- (2,{2+\FigGamma}) -- (-2,{-2+\FigGamma}) -- cycle;
  \draw[#2,line width=0.5pt] (-2,{-2-\FigGamma}) -- (2,{2-\FigGamma})
    (-2,{-2+\FigGamma}) -- (2,{2+\FigGamma});
  \end{scope}}
\newcommand{\PlaneAxes}{%
  \draw[axis] (-1.2,0) -- (1.25,0) node[right,lab] {$x_i$};
  \draw[axis] (0,-1.2) -- (0,1.25) node[above,lab] {$y_i$};
  \draw[cIn,line width=0.7pt] (-1,-1) rectangle (1,1);}
\newcommand{\PlaneStair}{(1,0) -- (1,1/3) -- (2/3,1/3) -- (2/3,2/3)
  -- (1/3,2/3) -- (1/3,1) -- (-1/3,1) -- (-1/3,2/3) -- (-2/3,2/3)
  -- (-2/3,1/3) -- (-1,1/3) -- (-1,-1/3) -- (-2/3,-1/3) -- (-2/3,-2/3)
  -- (-1/3,-2/3) -- (-1/3,-1) -- (1/3,-1) -- (1/3,-2/3) -- (2/3,-2/3)
  -- (2/3,-1/3) -- (1,-1/3) -- cycle}
\newcommand{\PlaneCross}{(0.5,-1) -- (0.5,-0.5) -- (1,-0.5) -- (1,0.5)
  -- (0.5,0.5) -- (0.5,1) -- (-0.5,1) -- (-0.5,0.5) -- (-1,0.5)
  -- (-1,-0.5) -- (-0.5,-0.5) -- (-0.5,-1) -- cycle}

\newcommand{\FigPlaneMidpoint}{%
\begin{tikzpicture}[paperfig,x=1.75cm,y=1.75cm]
\PlaneBox
\pgfmathsetmacro{\pa}{(1+\FigGamma)/2}
\pgfmathsetmacro{\pb}{(1-\FigGamma)/2}
\PlaneSquare
\PlaneStrip{cDiff!10}{cDiff!45}
\begin{scope}
\clip (-1,-1) rectangle (1,1);
\draw[cMid!75,line width=0.5pt] (-1,2) -- (2,-1) (-2,1) -- (1,-2);
\end{scope}
\fill[cMid!28] (\pa,\pb) -- (\pb,\pa) -- (-\pa,-\pb) -- (-\pb,-\pa) -- cycle;
\draw[cMid,line width=0.9pt]
  (\pa,\pb) -- (\pb,\pa) -- (-\pa,-\pb) -- (-\pb,-\pa) -- cycle;
\PlaneAxes
\node[lab,below] at (1,-1) {$R$};
\node[lab,below] at (-1,-1) {$-R$};
\node[lab,left] at (-1,1) {$R$};
\draw[cDiff,line width=1.5pt] (-0.5,-0.5) -- (0.5,0.5);
\draw[cDiff,line width=1.5pt] (-0.25,-0.75) -- (0.75,0.25);
\node[lab,text=cDiff!65!black,rotate=45,anchor=south] at (0.27,0.27) {$z_i=0$};
\node[lab,text=cDiff!65!black,rotate=45,anchor=north] at (0.27,-0.23) {$z_i=R/2$};
\node[lab,text=cMid,rotate=-45,anchor=south] at (0.5,0.5) {$x_i+y_i=R$};
\node[lab,text=cMid,rotate=-45,anchor=north] at (-0.5,-0.5) {$x_i+y_i=-R$};
\end{tikzpicture}}

\newcommand{\FigPlaneTernary}{%
\begin{tikzpicture}[paperfig,x=1.75cm,y=1.75cm]
\PlaneBox
\pgfmathsetmacro{\pa}{(1+\FigGamma)/2}
\pgfmathsetmacro{\pb}{(1-\FigGamma)/2}
\PlaneSquare
\fill[cCov!20] \PlaneStair;
\foreach \ft in {-0.66667,-0.33333,0.33333,0.66667}
  \draw[cIn!50,line width=0.3pt] (\ft,-1) -- (\ft,1) (-1,\ft) -- (1,\ft);
\draw[cCov!80!black,line width=1pt] \PlaneStair;
\fill[cMid!28] (\pa,\pb) -- (\pb,\pa) -- (-\pa,-\pb) -- (-\pb,-\pa) -- cycle;
\draw[cMid,line width=0.9pt]
  (\pa,\pb) -- (\pb,\pa) -- (-\pa,-\pb) -- (-\pb,-\pa) -- cycle;
\draw[cMid,densely dashed,line width=0.6pt]
  (1,0) -- (0,1) -- (-1,0) -- (0,-1) -- cycle;
\PlaneAxes
\foreach \ft/\fl in {-0.83333/2,-0.5/1,-0.16667/0,0.16667/0,0.5/1,0.83333/2} {
  \node[lab,below] at (\ft,-1) {$\fl$};
  \node[lab,left] at (-1,\ft) {$\fl$};}
\node[lab,anchor=north west] at (1.04,-1) {$b_i(x)$};
\node[lab,anchor=south] at (-1.08,1.04) {$b_i(y)$};
\foreach \ft/\fl in {0.33333/{\tfrac{R}{3}},0.66667/{\tfrac{2R}{3}},1/{R}}
  \node[lab,above] at (\ft,1) {$\fl$};
\end{tikzpicture}}

\newcommand{\FigPlaneBinary}{%
\begin{tikzpicture}[paperfig,x=1.75cm,y=1.75cm]
\PlaneBox
\PlaneSquare
\foreach \fa in {1,-1} \foreach \fb in {1,-1}
  \fill[pattern=north east lines,pattern color=cIn!50]
    ({\fa*0.5},{\fb*0.5}) rectangle (\fa,\fb);
\fill[cCov!20] \PlaneCross;
\draw[cCov!80!black,line width=1pt] \PlaneCross;
\fill[cMid!28] (1,0) -- (0,1) -- (-1,0) -- (0,-1) -- cycle;
\draw[cMid,line width=0.9pt] (1,0) -- (0,1) -- (-1,0) -- (0,-1) -- cycle;
\PlaneAxes
\draw[cDiff,line width=1.5pt] (-0.15,-0.85) -- (0.85,0.15);
\node[lab,text=cDiff!65!black,rotate=45,anchor=north] at (0.42,-0.28)
  {$x_i-y_i=v_i$};
\foreach \ft/\fl in {-1/{-\lambda},-0.5/{-\tfrac{\lambda}{2}},
    0.5/{\tfrac{\lambda}{2}},1/{\lambda}}
  \node[lab,below] at (\ft,-1) {$\fl$};
\node[lab,left] at (-1,1) {$\lambda$};
\node[lab,left] at (-1,0.5) {$\tfrac{\lambda}{2}$};
\end{tikzpicture}}

\newcommand{\FigPackCover}{%
\begin{tikzpicture}[paperfig,x=1.45cm,y=0.95cm]
\useasboundingbox (-1.75,-1.85) rectangle (1.62,1.8);
\draw[axis] (-1.5,0) -- (1.52,0);
\draw[cMid,line width=.9pt] ({-4/3},0) -- (0,{4/3}) -- ({4/3},0);
\node[lab,text=cMid] at (.92,1.0) {$f_r$};
\node[lab,below=1pt] at ({-4/3},0) {$-r$};
\node[lab,below=1pt] at ({4/3},0) {$r$};
\foreach \fx in {-.8,-.4,-.1,.4,.8}{
  \pgfmathsetmacro{\fw}{4/3-abs(\fx)}
  \draw[cMid!65,line width=.7pt] (\fx,0) -- (\fx,\fw);
  \fill[cMid] (\fx,\fw) circle (1.6pt);}
\draw[cDiff,line width=.8pt] (.8,{8/15}) circle (3pt);
\draw[cDiff,densely dashed,line width=.5pt] (.8,{8/15}) -- (.8,-1.5);
\foreach \fa/\fy in {-.8/-.5,-.4/-.83}{
  \draw[cIn!55,line width=1.2pt] (\fa,\fy) -- ({\fa+1},\fy);
  \fill[cIn!55] (\fa,\fy) circle (1.4pt);
  \draw[cIn!55,fill=white] ({\fa+1},\fy) circle (1.4pt);}
\foreach \fa/\fy/\fl in {-.1/-1.17/{I'},.4/-1.5/{I}}{
  \draw[cCov,line width=1.5pt] (\fa,\fy) -- ({\fa+1},\fy);
  \fill[cCov] (\fa,\fy) circle (1.5pt);
  \draw[cCov,fill=white] ({\fa+1},\fy) circle (1.5pt);
  \node[lab,text=cCov!70!black,anchor=east] at ({\fa-.06},\fy) {$\fl$};}
\end{tikzpicture}}

\newcommand{\FigPackBoxes}{%
\begin{tikzpicture}[paperfig,x=1.1cm,y=1.1cm]
\useasboundingbox (-2.2,-2.2) rectangle (1.95,1.8);
\draw[axis] (-1.15,-1.25) -- (1.72,-1.25) node[right,lab] {$x_1$};
\draw[axis] (-1.15,-1.25) -- (-1.15,1.43) node[above,lab] {$x_2$};
\draw[cCov!45,line width=1.1pt] (-.1,-1.86) -- (.9,-1.86);
\fill[cCov!45] (-.1,-1.86) circle (1.2pt);
\draw[cCov!45,fill=white] (.9,-1.86) circle (1.2pt);
\draw[cCov!45,line width=1.1pt] (-1.86,0) -- (-1.86,1);
\fill[cCov!45] (-1.86,0) circle (1.2pt);
\draw[cCov!45,fill=white] (-1.86,1) circle (1.2pt);
\draw[cCov,line width=1.5pt] (.4,-1.58) -- (1.4,-1.58);
\fill[cCov] (.4,-1.58) circle (1.5pt);
\draw[cCov,fill=white] (1.4,-1.58) circle (1.5pt);
\node[lab,text=cCov!70!black,anchor=west] at (1.47,-1.58) {$I$};
\draw[cCov,line width=1.5pt] (-1.58,-.8) -- (-1.58,.2);
\fill[cCov] (-1.58,-.8) circle (1.5pt);
\draw[cCov,fill=white] (-1.58,.2) circle (1.5pt);
\node[lab,text=cCov!70!black,above=2pt] at (-1.58,.2) {$J$};
\foreach \fa/\fb in {-.1/-.8,-.1/0,.4/0}
  \draw[cCov!45,line width=.5pt] (\fa,\fb) rectangle ({\fa+1},{\fb+1});
\fill[cCov!20] (.4,-.8) rectangle (1.4,.2);
\draw[cCov!80!black,line width=1pt] (.4,.2) -- (.4,-.8) -- (1.4,-.8);
\draw[cCov!80!black,densely dashed,line width=1pt]
  (.4,.2) -- (1.4,.2) -- (1.4,-.8);
\node[lab,text=cCov!60!black] at (.98,-.52) {$d_Id_J$};
\draw[cDiff!80,densely dashed,line width=.5pt]
  (.8,0) -- (.8,-1.58) (.8,0) -- (-1.58,0);
\foreach \fx/\fy in {-.8/-.8,-.4/.25,.4/-1.1,.8/0,-.1/1.3}
  \fill[cMid] (\fx,\fy) circle (2pt);
\draw[cDiff,line width=.9pt] (.8,0) circle (3.4pt);
\end{tikzpicture}}

\newcommand{\FigPackRatio}{%
\begin{tikzpicture}[paperfig,x=1.55cm,y=5.6cm]
\useasboundingbox (0.2,-0.1) rectangle (3.48,0.56);
\fill[cIn!14] (0.5,0) rectangle (0.88889,0.5);
\draw[axis] (0.5,0) -- (3.3,0) node[right,lab] {$r$};
\draw[axis] (0.5,0) -- (0.5,0.53);
\node[lab,anchor=west] at (1.0,0.44) {$A(r)/r^2$};
\draw[tick] (0.52,0) -- (0.48,0) node[left,lab] {$1$};
\foreach \ft in {2,3}
  \draw[tick] (\ft,0.008) -- (\ft,-0.008) node[below,lab] {$\ft$};
\draw[tick] (1,0.008) -- (1,-0.008);
\draw[cMid,line width=1pt] plot[domain=0.675:1,samples=40] (\x,{1/\x-1});
\draw[cMid,line width=1pt] plot[domain=1:2,samples=60]
  (\x,{(\x-1)*(2-\x)/(\x*\x)});
\draw[cMid,line width=1pt] plot[domain=2:3,samples=60]
  (\x,{(\x-2)*(3-\x)/(\x*\x)});
\draw[cMid,line width=1pt] plot[domain=3:3.22,samples=20]
  (\x,{(\x-3)*(4-\x)/(\x*\x)});
\draw[cDiff,densely dashed,line width=0.7pt] (0.5,0.125) -- (3.25,0.125);
\draw[tick] (0.52,0.125) -- (0.48,0.125)
  node[left,lab,text=cDiff!70!black] {$\tfrac98$};
\foreach \ft in {0.88889,1.33333} {
  \draw[cDiff!70,densely dotted,line width=0.6pt] (\ft,0.125) -- (\ft,0);
  \fill[cDiff] (\ft,0.125) circle (1.8pt);}
\node[lab,below,text=cDiff!70!black] at (0.88889,0) {$\tfrac89$};
\node[lab,below,text=cDiff!70!black] at (1.33333,0) {$\tfrac43$};
\end{tikzpicture}}

\newcommand{\FigWinSections}{%
\begin{tikzpicture}[paperfig,x=1.5cm,y=1.5cm]
\useasboundingbox (-1.42,-1.36) rectangle (1.5,1.46);
\PlaneSquare
\fill[cCov!20] \PlaneCross;
\draw[cCov!80!black,line width=.8pt] \PlaneCross;
\PlaneAxes
\foreach \ft in {-1,1}
  \node[lab,below] at (\ft,-1) {$\ft$};
\node[lab,left] at (-1,1) {$1$};
\foreach \fa/\fb/\fx/\fy/\fl/\fu/\fv in
  {-1/0/0/1/{w=-1}/-.52/.52,-.5/-.5/.5/.5/{w=0}/-.28/-.24,0/-1/1/0/{w=1}/.5/-.46}{
  \draw[cMid,line width=1.5pt] (\fa,\fb) -- (\fx,\fy);
  \fill[cMid] (\fa,\fb) circle (1.8pt);
  \draw[cMid,fill=white,line width=.7pt] (\fx,\fy) circle (1.8pt);
  \node[lab,text=cMid,rotate=45,anchor=south] at (\fu,\fv) {$\fl$};}
\end{tikzpicture}}

\newcommand{\FigWinWeight}{%
\begin{tikzpicture}[paperfig,x=1.55cm,y=.8cm]
\useasboundingbox (-1.52,-1.85) rectangle (1.68,2.95);
\fill[cDiff!14] (-1,0) rectangle (-.5,2);
\fill[cIn!14] (-.5,0) rectangle (.5,1);
\fill[cDiff!14] (.5,0) rectangle (1,2);
\fill[cCov!45] (.25,0) rectangle (.5,1);
\fill[cCov!45] (.5,0) rectangle (1,2);
\draw[axis] (-1.4,0) -- (1.5,0) node[right,lab] {$t$};
\draw[cDiff,line width=1.1pt] (-1,2) -- (-.5,2) (.5,2) -- (1,2);
\draw[cMid,line width=1.1pt] (-.5,1) -- (.5,1);
\node[lab,text=cDiff!70!black] at (-.75,2.35) {$2$};
\node[lab,text=cDiff!70!black] at (.75,2.35) {$2$};
\node[lab,text=cMid] at (0,1.35) {$1$};
\node[lab,text=cIn!80!black] at (-1.25,1.6) {$g$};
\foreach \fx/\fl in {-1/{-\lambda},-.5/{-\tfrac{\lambda}{2}},
    .5/{\tfrac{\lambda}{2}},1/{\lambda}}
  \node[lab,below=1pt] at (\fx,0) {$\fl$};
\draw[cCov!80!black,densely dashed,line width=.5pt] (.25,-1.05) -- (.25,2.5);
\draw[cCov!80!black,densely dashed,line width=.5pt] (1.25,-1.05) -- (1.25,2.5);
\draw[cCov,line width=1.4pt] (.25,-1.2) -- (1.25,-1.2);
\fill[cCov] (.25,-1.2) circle (1.6pt);
\draw[cCov,fill=white,line width=.7pt] (1.25,-1.2) circle (1.6pt);
\node[lab,below=2pt,text=cCov!70!black] at (.75,-1.2) {$B_0-w$};
\end{tikzpicture}}

\newcommand{\FigWinDisjoint}{%
\begin{tikzpicture}[paperfig,x=1.4cm,y=1.4cm]
\useasboundingbox (-1.5,-1.38) rectangle (1.5,1.44);
\fill[cDiff!20] (-1,-1) rectangle (1,1);
\fill[pattern=north east lines,pattern color=cDiff!70] (-1,-1) rectangle (1,1);
\begin{scope}
\clip (-1,-1) rectangle (1,1);
\foreach \fa in {-1,0,1}\foreach \fb in {-1,0,1}{
  \pgfmathsetmacro{\fy}{.25*\fa+1.25*\fb}
  \fill[cCov!16!white] ({\fa-.5},{\fy-.5}) rectangle ({\fa+.5},{\fy+.5});}
\end{scope}
\begin{scope}
\clip (-1.43,-1.32) rectangle (1.43,1.32);
\foreach \fa in {-1,0,1}\foreach \fb in {-2,-1,0,1,2}{
  \pgfmathsetmacro{\fy}{.25*\fa+1.25*\fb}
  \draw[cCov!85!black,line width=.7pt]
    ({\fa-.5},{\fy+.5}) -- ({\fa-.5},{\fy-.5}) -- ({\fa+.5},{\fy-.5});
  \draw[cCov!85!black,densely dashed,line width=.6pt]
    ({\fa-.5},{\fy+.5}) -- ({\fa+.5},{\fy+.5}) -- ({\fa+.5},{\fy-.5});
  \pgfmathtruncatemacro{\fvis}{abs(\fy)<1.28 ? 1 : 0}
  \ifnum\fvis=1 \fill[cMid] (\fa,\fy) circle (1.8pt); \fi}
\end{scope}
\draw[cDiff!85!black,line width=.8pt] (-1,-1) rectangle (1,1);
\node[lab,fill=white,inner sep=1pt,text=cDiff!70!black] (gap) at (.75,1.2) {gap};
\draw[->,cDiff!85!black,line width=.7pt] (gap.south west) -- (.18,.63);
\end{tikzpicture}}
\title{$\ell_{\infty}$ norm SVP in $\widetilde{O}(3^n)$}
\author{Yiming Gao\\[0.4em]
\small School of Cyber Science and Technology,\\
\small University of Science and Technology of China\\[0.2em]
\small \texttt{qw1234567@mail.ustc.edu.cn}}
\date{}

\begin{document}
\maketitle
\vspace{-1.5em}
\begin{abstract}
We give a classical randomized algorithm for the Shortest Vector Problem in the infinity norm in $\OO(3^n)$ time and space, improving the previous bound of $2^{2.443n+o(n)}$. A variant of the algorithm lists all shortest nonzero vectors within the same bounds. Both results, including their time and space bounds, are formally verified in Lean~4.

We use the counting moment framework of Gaussian difference dynamics to construct lists with equal expected counts at each lattice point in successively smaller cubes. The key is to restrict the midpoint of each pair to a fixed inner cube, so that every allowed difference has the same region of possible midpoints. Averaging random lattice translates then preserves these expected counts through repeated differencing. We control the dependence between entries through bounds on the second counting moment, using a weighted packing estimate for cubes to bound the variation in midpoint counts. Finally, two searches based on coordinate comparisons generate the intermediate lists and recover a shortest lattice vector within the claimed time bound.
\end{abstract}

\section{Introduction}\label{sec:intro}
Let $L=B\Z^n\subset\R^n$ be a full-rank lattice. The Shortest Vector Problem in the infinity norm asks for a nonzero vector $v\in L$ satisfying
\[
 \infnorm v=\lambda_1^{(\infty)}(L)
 :=\min_{x\in L\setminus\{0\}}\max_{1\le i\le n}|x_i|.
\]
Equivalently, it asks for the smallest cube centered at the origin that contains a nonzero lattice point.
The infinity norm naturally describes coordinatewise error bounds in
cryptanalytic lattice problems, including the Hidden Number
Problem~\cite{BV96} and attacks on DSA and ECDSA with partially known
nonces~\cite{HGS01,NS03}.

Previous provable algorithms for exact $\ell_\infty$-SVP are linear
sieves adapted to the product geometry of the cube~\cite{AM18,Muk21},
and the best of them runs in $2^{2.443n+o(n)}$ time and
space~\cite{Muk21}.

In the Euclidean setting, Gao, Han, and Hu~\cite{GDD} introduced
Gaussian difference dynamics (GDD), a counting framework for repeated
differencing. It yields classical randomized algorithms for discrete
Gaussian sampling at widths a fixed factor above the smoothing parameter
in $(3/2)^{n/2+o(n)}$ time, with exponent approximately $0.292$, and for
exact Euclidean SVP in $2^{0.413n+o(n)}$ time, both on arbitrary lattices.
The core idea is to maintain bounds on the first two counting moments
throughout the list iteration. The first describes the expected number
of entries at each lattice point relative to a Gaussian reference, while
the second counts ordered pairs of distinct entries and limits their
clustering. Conditional independence between separately generated child
lists gives a closed recursion for these moments even though entries
within either list may share inputs and ancestors. The resulting bounds
track the supply of useful vectors through successive rounds and connect
expected target counts to the probability of success.

The moment recursion itself is general, but the efficient Euclidean
constructions rely on Gaussian completion of squares, thin spherical
shells, and geometric searches for short differences. These ingredients
both preserve the Gaussian reference through the iteration and determine
the cost of generating the next list. For the infinity norm, we instead
seek equal expected counts at the lattice points of each successive
cube. Directly restricting a difference to a smaller cube does not
preserve this reference: its region of possible midpoints varies with
the difference vector and shrinks toward the boundary. Controlling the
first two counting moments therefore calls for a cube transition with
suitable representation counts and variance bounds, together with
searches adapted to that transition.

We give a classical randomized algorithm for exact $\ell_\infty$-SVP in
$\OO(3^n)$ time and space, reducing the exponent from $2.443$ to
$\log_2 3\approx1.585$. Restricting every retained pair to the same
midpoint cube makes the midpoint volume independent of the allowed
difference, and averaging a random lattice translate gives exact
first-moment closure. A weighted packing bound with factor $(9/8)^n$,
combined with a random superlattice, controls the variation in midpoint
counts and determines the required list size. Finally, ternary coordinate
classes generate the intermediate lists, while a binary condition
combined with quotient labels recovers a shortest vector. The candidate
and report bounds put both searches within $\OO(3^n)$ time. Throughout
the paper, $\OO$ suppresses factors polynomial in the dimension and the
binary input length.

The base $3$ also has a geometric interpretation: the maximum lattice
kissing number in the infinity norm is $3^n-1$~\cite[Equation~(1.3)]{Henk02},
attained by $\Z^n$ through the shortest vectors
$\{-1,0,1\}^n\setminus\{0\}$. Listing all shortest vectors therefore
needs $3^n-1$ outputs in the worst case, and a variant of our algorithm
matches this (\cref{thm:all-shortest}). Our terminal search is
controlled by a related weighted packing bound that is also attained by
the integer lattice, as discussed in \cref{sec:conclusion}.

\subsection{Our results}
\begin{theorem}[Main theorem]\label{thm:main}
There is a classical randomized algorithm that, given a nonsingular rational basis of a full-rank lattice $L\subset\R^n$, always returns a nonzero vector of $L$, which is shortest in the infinity norm with probability at least $2/3$. Every execution uses
\[
 T=\OO(3^n)\qquad\text{and}\qquad S=\OO(3^n).
\]
\end{theorem}

\begin{theorem}[Enumerating all shortest vectors]\label{thm:all-shortest}
There is a classical randomized algorithm that, given a nonsingular rational basis of a full-rank lattice $L\subset\R^n$, always returns a nonempty list of distinct nonzero vectors of $L$, which consists of all shortest nonzero vectors of $L$ in the infinity norm with probability at least $2/3$. Every execution uses
\[
 T=\OO(3^n)\qquad\text{and}\qquad S=\OO(3^n).
\]
\end{theorem}

\paragraph{Application to Subset Balancing.}
Combining \cref{thm:main} with the reduction of Gao, Feng, Hu, and
Pan~\cite[Theorem~3.3]{GFHP26} to one $(n+1)$-dimensional
$\ell_\infty$-SVP instance gives the following consequence.

\begin{corollary}\label{cor:subset-balancing}
Given $x\in\Z^n$ and an integer $d\ge1$, there is a randomized algorithm
that, with probability at least $2/3$, finds a vector
$c\in\{-d,\ldots,d\}^n\setminus\{0\}$ satisfying $c\cdot x=0$, or
correctly reports that no such vector exists. It uses $\OO(3^n)$ time
and space.
\end{corollary}

For every fixed $d\ge5$, this is exponentially faster than the standard
meet-in-the-middle bound $\OO((2d+1)^{n/2})$, improving the threshold
$d\ge15$ obtained in~\cite{GFHP26}. In the same range, the exponent
$\log_2 3\approx1.585$ is smaller than the optimized bound from the
analysis of Randolph and W\k{e}grzycki~\cite{RW26}, which is
approximately $1.609$ at $d=5$. This gives the fastest known classical
algorithm for Subset Balancing in the worst case for these $d$.

\paragraph{Formal verification.}
We verified both \cref{thm:main,thm:all-shortest} in Lean~4 for fixed
programs on a randomized log-cost RAM.\footnote{Lean~4 source code and
documentation: \LeanRepoURL.} Compiling these programs to a randomized
word RAM with $w=\Theta(n+\log\bits(B))$-bit words preserves their
output distributions and their $\OO(3^n)$ time and space bounds. The
formalization uses only Lean's three standard foundational axioms and
introduces no unproven assumptions.

\subsection{Related work}
Ajtai, Kumar, and Sivakumar introduced randomized sieving for lattice
problems~\cite{AKS01}. Bl\"omer and Naewe~\cite{BN09} extended this
approach to general $\ell_p$ norms, obtaining single-exponential
algorithms for exact SVP and approximation algorithms for CVP.
Dadush, Peikert, and Vempala obtained single-exponential algorithms for
SVP in arbitrary norms by using $M$-ellipsoid coverings~\cite{DPV11}.

For the infinity norm, Aggarwal and Mukhopadhyay~\cite{AM18}
specialized the AKS framework to the product geometry of the cube.
A partition of the cube into smaller cubes contains at most one center
in each part, so the center of a vector is found from the intervals of
its coordinates and the sieve becomes linear in the list size. Together
with a birthday argument, this gives exact $\ell_\infty$-SVP in
$2^{2.82n+o(n)}$ time and space. They also gave approximation
algorithms for SVP and CVP in the infinity norm, with the exponential
base for approximate SVP approaching $3$ as the approximation factor
grows. Mukhopadhyay~\cite{Muk21} extended the linear sieve to every
$\ell_p$ norm, with a general SVP bound of $2^{2.751n+o(n)}$ time and
space and approximation algorithms for SVP and CVP in these norms, and
for $p=\infty$ combined this analysis with the cube partition
of~\cite{AM18} to reach $2^{2.443n+o(n)}$. Our algorithm solves the
exact problem with base $3$, which these approximation algorithms
approach only as the approximation factor grows.

Geometric coverings also give approximation algorithms for CVP.
Eisenbrand, H\"ahnle, and Niemeier~\cite{EHN11} used cube coverings to
obtain a randomized $(1+\epsilon)$-approximation algorithm for
$\mathrm{CVP}_\infty$ in $2^{O(n)}(\log(1/\epsilon))^{O(n)}$ time.

Random superlattices and quotient labels appear in the Euclidean
algorithms of Gao, Feng, and Hu~\cite{GFH26a,GFH26b}. A superlattice
contains every target vector of the input lattice, and equal labels
certify that a difference lies in the input lattice.

\paragraph{Organization.}
\Cref{sec:overview} explains the construction and derives the main cost balance. \Cref{sec:prelim} fixes notation and the standard lattice tools. \Cref{sec:cubes,sec:super} control cube counts on a random superlattice. \Cref{sec:lists} builds the lists and proves their counting bounds. \Cref{sec:internal,sec:terminal} give the two pair searches. \Cref{sec:algorithm} assembles the algorithms and proves \cref{thm:main,thm:all-shortest}. \Cref{sec:conclusion} discusses the connection with lattice kissing numbers. Appendix~\ref{app:basis} proves the basis preprocessing lemma, and Appendix~\ref{app:implementation} gives the finite arithmetic details.

The author used generative AI to assist with algorithm design, mathematical analysis, manuscript drafting, and formal verification, as described in the \hyperref[sec:ai-methodology]{AI methodology statement} at the end of the paper.

\section{Technique overview}\label{sec:overview}
Fix a shortest vector $v\in L\setminus\{0\}$ for the analysis, and write $\lambda=\infnorm v$. There are two tasks: make $v$ appear as a difference between two lists, and find that difference within the claimed time. We first explain the construction assuming that $\lambda$ is known, then remove this assumption by trying polynomially many scales.

The lists are generated using the counting recursion of GDD~\cite{GDD}.
At each level, we preserve equal expected counts at the allowed lattice
points and bound the expected number of ordered pairs of distinct
entries. These two bounds track the available supply and control
clustering, allowing us to turn a large expected number of target
representations into a high success probability.

\subsection{Preserving uniform counts by midpoint restriction}
We seek a difference rule that preserves equal expected counts across
each successive cube.
Write
\[
 Q_R=[-R,R)^n,\qquad B_R=[-R/2,R/2)^n.
\]
The input lists lie in $Q_R$, while $B_R$ is the allowed midpoint
region. For a fixed $\gamma<1$, keep pairs from the two lists satisfying
\begin{equation}\label{eq:overview-rule}
 z=x-y\in Q_{\gamma R},\qquad M=(x+y)/2\in B_R.
\end{equation}
For every allowed $z$, the complete set of possible midpoints is exactly $B_R$. Conversely, $M\in B_R$ and $z\in Q_{\gamma R}$ imply $M\pm z/2\in Q_R$. The volume of the midpoint set is therefore $R^n$, independently of $z$.

\begin{figure}[htbp]
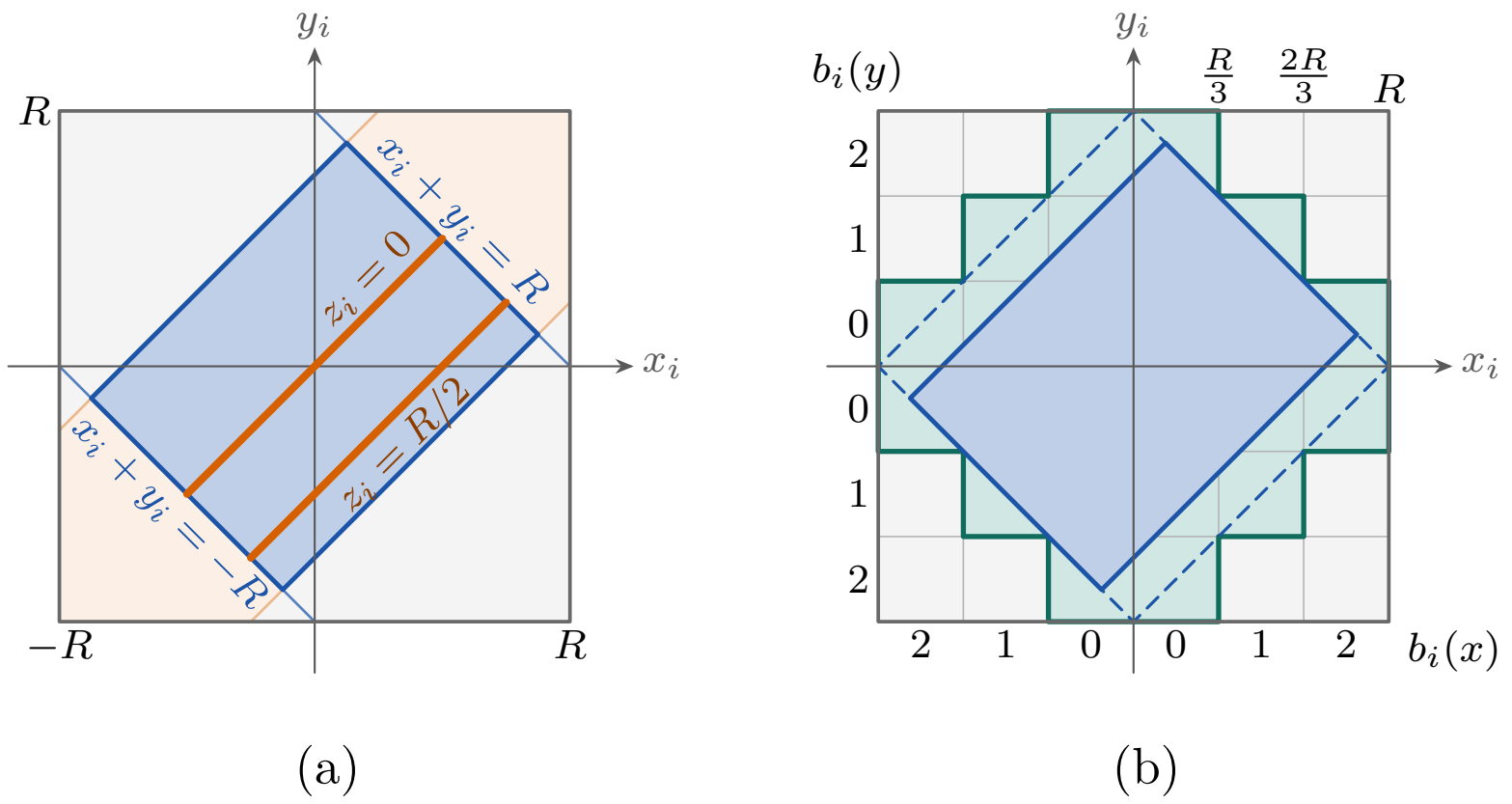

\centering
\begin{minipage}[b]{.32\textwidth}\centering
\FigPlaneMidpoint\\[1pt]
{\small(a)}
\end{minipage}\hfill
\begin{minipage}[b]{.32\textwidth}\centering
\FigPlaneTernary\\[1pt]
{\small(b)}
\end{minipage}\hfill
\begin{minipage}[b]{.32\textwidth}\centering
\FigPlaneBinary\\[1pt]
{\small(c)}
\end{minipage}
\caption{The pair conditions in one coordinate, drawn with
$\gamma=3/4$ in (a,b) and $R_0=\ell=\lambda$ in (c). (a) Every allowed
difference has the same midpoint interval. (b) Six ternary class pairs
cover the dashed diamond $|x_i|+|y_i|\le R$. (c) The binary cross
contains the target diamond.}\label{fig:plane}
\end{figure}

Restricting only $x-y$ would leave a midpoint interval of length
$2R-|z_i|$ in coordinate $i$, so the continuous output density would be
proportional to $\prod_{i=1}^n(2R-|z_i|)$ and would decay exponentially
toward the corners of $Q_{\gamma R}$. The midpoint condition in
\eqref{eq:overview-rule} cuts every such interval to the same length $R$
(\cref{fig:plane}(a)), giving a constant output density on
$Q_{\gamma R}$ in the continuous model.

For lattice points, the number of admissible midpoints also depends on
the lattice translate. We build the lists along a binary tree, choosing
a new random translate at each node and generating the two child lists
with independent randomness conditional on their translates. Averaging
the translate makes the expected output count equal at every lattice
point of the smaller cube. The second counting moment accounts for both
the variation in midpoint counts and the dependence caused by pairs
that share an input entry. Its recursion requires a bound on the
midpoint count's variance, which we obtain by passing to a random
superlattice.

\subsection{Controlling midpoint variance with a random superlattice}
A cube near the shortest vector scale can contain very different
numbers of points in different translates of the input lattice.
For a lattice $\Lambda$ and a uniform translate $t+\Lambda$, let
\[
 J_R^\Lambda(t)=\#\bigl((t+\Lambda)\cap B_R\bigr).
\]
We measure this variation by
\begin{equation}\label{eq:overview-theta}
 \Theta_R(\Lambda^*)
 =\frac{\Var(J_R^\Lambda)}{\E[J_R^\Lambda]^2},
 \qquad
 \E[J_R^\Lambda]=\frac{R^n}{\det\Lambda}.
\end{equation}
We work in a random superlattice $\Gamma\supset L$, of determinant
$D$, chosen so that these relative variances are small at all widths
used by the construction.

The main geometric estimate is
\begin{equation}\label{eq:overview-packing}
 1+\Theta_R(L^*)\le
 C^n\frac{\det L}{\lambda^n},
 \qquad C=\frac98,\qquad R\ge\frac89\lambda.
\end{equation}
The second moment of $J_R^L$ counts ordered pairs of lattice points in the translated cube. Grouping these pairs by their difference $w\in L$ turns it into a sum of the overlap volumes $\prod_i(R-|w_i|)_+$ of $B_R$ and $B_R+w$, and since distinct points of $L$ are at distance at least $\lambda$ in the infinity norm, bounding this sum is a weighted packing problem for cubes.

The constant can already be seen in one dimension. An interval of length $4/3$, translated uniformly modulo $\Z$, contains one point with probability $2/3$ and two points with probability $1/3$, so its second moment divided by the square of its mean is $9/8$, and products of these intervals give $(9/8)^n$ on $\Z^n$. A general lattice is not a product, and its coordinate projections can be crowded even though its points are separated. We therefore cover the weights in each coordinate by weighted unit intervals and multiply these covers into unit cubes, each containing at most one point of $L/\lambda$.

Passing to a random superlattice of prime index $p$ reduces the
expected relative variance to at most $\Theta_R(L^*)/p$, plus an
error term controlled in \cref{sec:super}. We choose
\[
 p\asymp C^n\frac{\det L}{\lambda^n}\poly(n),
 \qquad
 D\asymp\lambda^n C^{-n}/\poly(n).
\]
This makes the relative variance small at all required widths with
high probability. The density also determines the list size: for
lists of a fixed average size, decreasing $D$ decreases the expected
number of representations of the target vector, as the following
calculation shows.

\subsection{Choosing the list size from target representations}
Suppose a final list has expected count $A$ at each point of a common translate $z+\Gamma$ inside $Q_\lambda$. If its average size is $N$, then
\[
 A=\frac{ND}{(2\lambda)^n}.
\]
Every midpoint in $B_\lambda$ represents $v$ through the endpoints $M+v/2$ and $M-v/2$. The number of available midpoints is about $\lambda^n/D$. Thus the expected number of representations in the two lists is
\begin{equation}\label{eq:overview-target}
 A^2\frac{\lambda^n}{D}
 =\frac{N^2D}{4^n\lambda^n}
 \asymp\frac{N^2}{4^n C^n\poly(n)}.
\end{equation}
At a fixed list size, this count is proportional to $D$. Halving $D$
doubles the number of midpoint locations, but each endpoint then has
half the expected count, so each location yields a quarter as many
pairs. The density required by the variance bound therefore sets the
list size, and \eqref{eq:overview-target} suggests
\[
 N=(2\sqrt C)^n\poly(n)
   =(3/\sqrt2)^n\poly(n).
\]
A suitable polynomial factor makes the target count large. To turn this expectation into a success probability, we also control pairs of distinct entries within each list. Repeated vector values are kept as separate entries, so that every retained pair contributes one new entry to the counting recursion.

At every round, the reference count per point is $A_R=ND/(2R)^n$. We thin each eligible pair with probability
\[
 q=\frac{(2/\gamma)^n}{N}.
\]
Then $qA_R^2(R^n/D)=A_{\gamma R}$, so the same reference list size $N$
is maintained throughout the tree. We choose a fixed
$\sqrt{8/9}<\gamma<1$, which ensures $q\le1$ while shortening the
radius by a constant factor.

The construction starts in a cube whose radius is polynomially larger
than $\lambda$, where we can sample the initial lists directly with
the required counting moments. The tree therefore has depth
$O(\log n)$ and polynomially many nodes, so its leading exponential
cost is determined by the list size and the pair searches at each node.

\subsection{Finding pairs with coordinatewise tests}
Gaussian difference dynamics~\cite{GDD} uses BDGL product
codes~\cite{BDGL16} to find pairs with short Euclidean differences.
In the infinity norm, the difference and midpoint conditions can be
tested coordinate by coordinate. We exploit this structure directly:
three classes per coordinate generate candidates for each recursive
merge, while a binary condition and quotient labels recover a shortest
vector from the two root lists.

For an internal pair, the midpoint restriction in
\eqref{eq:overview-rule} has a second effect. Together with
$x-y\in Q_{\gamma R}$ it gives
\[
 |x_i|+|y_i|=\max\{|x_i-y_i|,|x_i+y_i|\}\le R,
\]
whereas restricting $x-y$ alone would allow $|x_i|+|y_i|$ close to $2R$.
Divide each absolute coordinate into three classes, according to the intervals $[0,R/3]$, $(R/3,2R/3]$, and $(2R/3,R]$. If the class numbers are $b_i(x),b_i(y)\in\{0,1,2\}$, every eligible pair satisfies
\[
 b_i(x)+b_i(y)\le2,
\]
as shown in \cref{fig:plane}(b). There are six compatible class pairs out of nine in each coordinate. A table indexed by $\{0,1,2\}^n$ reports the corresponding pairs without visiting empty combinations. The candidate count has scale
\[
 N^2(2/3)^n=3^n\poly(n).
\]
Read in the other direction, a merge of cost $3^n$ can afford $N^2$ up
to $3^n(3/2)^n=(9/2)^n$, which is exactly the $4^n(9/8)^n$ required by
\eqref{eq:overview-target}. The ternary filter and the packing constant
balance with no slack. These tables also account for the $\OO(3^n)$
space bound, which exceeds the space needed to store the lists.

At the root, the task is to recover a vector of $L$ from the two lists.
Equal quotient labels recognize differences in $L$, but on a lattice as
dense as $\Z^n$ they still leave $N^2/p=4^n\poly(n)$ pairs. The search
therefore needs a coordinate condition as well, with tables small
enough to build inside a single label group. We use a binary rule: in
every coordinate, at least one endpoint must lie in
$[-\lambda/2,\lambda/2)$, as in \cref{fig:plane}(c). This keeps every
representation of a shortest vector whose midpoint lies in
$B_\lambda$, since $|v_i|\le\lambda$ ensures that at least one of
$M_i+v_i/2$ and $M_i-v_i/2$ lies in the inner interval.

The binary rule also lets us use the separation of $L$ to bound the
number of reports. The two lists lie in a common translate of the
denser lattice $\Gamma$, so their differences lie in $\Gamma$, and
equal labels restrict them further to $L$, where distinct points are at
distance at least $\lambda$. In one dimension, inclusion and
exclusion rewrites
the compatibility count as the overlap of the inner interval with a
nonnegative weight. This weight is one on the inner interval and two
on the remainder of $[-\lambda,\lambda)$, so its total mass is
$\lambda+2\lambda=3\lambda$. Taking products over coordinates gives a
weighted count in translates of $B_\lambda$. Each such translate
contains at most one point of $L$, which bounds the lattice sum by the
total weight $(3\lambda)^n$. Thus, if $P$ counts pairs satisfying both
conditions, then
\[
 \E[P]\le\frac{A^2(3\lambda)^n}{D}=3^n\poly(n).
\]
This bound is exactly $3^n$ times the target count in
\eqref{eq:overview-target}, for every choice of $N$ and $D$.

To find these pairs, we check small label groups directly and use a
binary prefix table for large groups. Splitting at group size $2^{n/2}$
gives a deterministic search that costs $\poly(n)2^{n/2}$ per list
entry and $\poly(n)$ per reported pair. Since each root list has
expected size $N$, the expected cost is
\[
 \poly(n)\bigl(N2^{n/2}+\E[P]\bigr)=3^n\poly(n).
\]
Like the candidate count of the internal searches, this is a bound on
expected work. The work limits of \cref{sec:algorithm} turn these
expectations into the bound on every execution in \cref{thm:main}.

\begin{table}[htbp]
\centering
\begin{tabular}{@{}lll@{}}
\toprule
Part of the algorithm & Work or storage & Exponential factor\\
\midrule
List generation & $N$ & $(3/\sqrt2)^n$\\
Internal candidate pairs & $N^2(2/3)^n$ & $3^n$\\
Final searches within label groups & $N2^{n/2}$ & $3^n$\\
Final pair reports & $A^2(3\lambda)^n/D$ & $3^n$\\
Ternary tables, the dominant storage & $3^n$ entries & $3^n$\\
\bottomrule
\end{tabular}
\caption{The cost balance. Polynomial factors are omitted in this overview.}\label{tab:balance}
\end{table}

\FloatBarrier
\section{Preliminaries}\label{sec:prelim}
We use $\log$ for the natural logarithm and $\log_2$ for the binary logarithm. For a real number $t$, write $(t)_+=\max\{t,0\}$. All lists are finite sequences of entries with distinct identifiers. Different entries may store the same vector.

\subsection{Lattices and basic tools}
For a nonsingular rational matrix $B$, let
\[
 L=B\Z^n,\qquad \det L=|\det B|,\qquad L^*=B^{-T}\Z^n.
\]
We write $\lambda_i^{(p)}(L)$ for the $i$th successive minimum in the $\ell_p$ norm. Unless a different lattice is specified, $\lambda$ denotes $\lambda_1^{(\infty)}(L)$. The inequalities
\[
 \infnorm x\le\eunorm x\le\sqrt n\infnorm x
\]
will be used for preprocessing only.

Minkowski's second theorem, applied to the cube, gives~\cite{Henk02}
\begin{equation}\label{eq:minkowski}
 \prod_{i=1}^n\lambda_i^{(\infty)}(L)\le\det L,
 \qquad
 \lambda_n^{(\infty)}(L)\le\frac{\det L}{\lambda^{n-1}}.
\end{equation}
In particular, $\det L\ge\lambda^n$. We also use the Euclidean transference bound~\cite{Ban93}
\begin{equation}\label{eq:transference}
 \lambda_n^{(2)}(\Lambda)\lambda_1^{(2)}(\Lambda^*)\le n.
\end{equation}

The following basis lemma places the starting cube within a polynomial
factor of the final scale. We use the general Euclidean preprocessing
result of~\cite[Proposition~3.2]{GDD}.

\begin{lemma}\label{lem:basis}
Put $\Ared(n)=2(n+1)$. From a rational basis of a rank-$n$ lattice $\Lambda$, one can compute a size-reduced basis $(b_1,\ldots,b_n)$ such that
\[
 \max_i\eunorm{b_i^*}\le\Ared(n)\lambda_n^{(2)}(\Lambda).
\]
The time and workspace are $2^{n+o(n)}\poly(\bits(B))$. All projected bases, lifted vectors, and integer coordinates have polynomial encoding length.
\end{lemma}
\begin{proof}
Taking $k=\lfloor n/2\rfloor$ in~\cite[Proposition~3.2]{GDD} gives
the claim for $n\ge4$, and the remaining dimensions use exact SVP.
Appendix~\ref{app:basis} gives the proof from~\cite{ALNS20,MV13}.
\end{proof}

Here $b_i^*$ are the Gram--Schmidt vectors. For the computed basis matrix $B_{\mathrm{red}}=(b_1,\ldots,b_n)$, size reduction gives
\begin{equation}\label{eq:cell-radius}
 \sup_{x\in B_{\mathrm{red}}[0,1)^n}\infnorm x
 \le\sum_i\eunorm{b_i}
 \le n^2\Ared(n)\lambda_n^{(2)}(\Lambda).
\end{equation}

\subsection{Boxes and finite shifts}
We use the half-open boxes
\[
 Q_R=[-R,R)^n,\qquad B_R=[-R/2,R/2)^n.
\]
Thus $Q_R$ has radius $R$ and $B_R$ has side length $R$. Half-open endpoints assign every grid point to exactly one interval when we partition a coordinate.

All random shifts are chosen from a finite quotient. For a fixed rational superlattice $\Gamma$, choose a rational $\delta>0$ such that
\begin{equation}\label{eq:grid-assumptions}
 \Gamma\subseteq G:=\delta\Z^n.
\end{equation}
Then $G/\Gamma$ has $D/\delta^n$ elements, where $D=\det\Gamma$. A uniform shift $a\in G/\Gamma$ satisfies
\begin{equation}\label{eq:one-membership}
 \Prb[x\in a+\Gamma]=\frac{\delta^n}{D}\qquad(x\in G).
\end{equation}
A basis in Hermite normal form permits uniform quotient sampling in polynomial time. We describe the exact arithmetic in Appendix~\ref{app:implementation}.

\subsection{Counts in random lists}
For a list $\List$, let $N_x$ be the number of entries equal to $x$. Define
\begin{equation}\label{eq:count-def}
 m_{\List}(x)=\E[N_x],\qquad
 S_{\List}(x,x')=\E\bigl[N_xN_{x'}-\ind_{x=x'}N_x\bigr].
\end{equation}
The second quantity counts ordered pairs of distinct entries. In particular, $S_{\List}(x,x)=\E[N_x(N_x-1)]$.

At a node with radius $R$ and prescribed coset $c+\Gamma$, the reference is
\begin{equation}\label{eq:reference}
 A_R=\frac{ND}{(2R)^n},\qquad
 m_{\List}(x)=A_R,\qquad
 S_{\List}(x,x')\le e^\varepsilon A_R^2
 \quad(x,x'\in(c+\Gamma)\cap Q_R).
\end{equation}
The moments in \eqref{eq:reference} average the node's private randomness and keep its coset fixed. The first equality gives the expected supply at each point, while the second inequality limits clustering. For example, if each of two independent lists consists of $N$ copies of a single uniform draw, every difference has the same expected number of pairs as for independent draws, yet all $N^2$ pairs test one and the same difference.

When a node chooses the two child cosets, we first fix those cosets and multiply the moments of the independent child subtrees, then average the common choice of child cosets. This order will be used in both the list analysis and the work estimates.

\section{Weighted cube counts}\label{sec:cubes}
The number of points in a translated cube is the geometric quantity needed by the list construction. We first express its variance as a nonnegative Fourier sum, then bound that sum by a packing argument in the original lattice.

\subsection{The variance of a translated box}
For a full-rank lattice $\Lambda$ and $R>0$, define
\begin{equation}\label{eq:theta-def}
 \Theta_R(\Lambda^*)=
 \sum_{\xi\in\Lambda^*\setminus\{0\}}
 \prod_{i=1}^n\sinc^2(R\xi_i).
\end{equation}
Here $\sinc(t)=\sin(\pi t)/(\pi t)$, with $\sinc(0)=1$.
By the following lemma, $\Theta_R(\Lambda^*)$ is the relative variance
of the translated box count used in \cref{sec:overview}.

\begin{lemma}\label{lem:parseval}
Let $t$ be uniform modulo $\Lambda$, and put
\[
 J_R(t)=\#\bigl((t+\Lambda)\cap B_R\bigr),
 \qquad \mu_R=R^n/\det\Lambda.
\]
Then
\begin{equation}\label{eq:parseval}
 \E[J_R]=\mu_R,\qquad
 \frac{\Var J_R}{\mu_R^2}=\Theta_R(\Lambda^*),
\end{equation}
and
\begin{equation}\label{eq:tent-poisson}
 1+\Theta_R(\Lambda^*)
 =\frac{\det\Lambda}{R^{2n}}
   \sum_{w\in\Lambda}\prod_{i=1}^n(R-|w_i|)_+.
\end{equation}
\end{lemma}
\begin{proof}
The periodization of $\ind_{B_R}$ is bounded on a fundamental domain and hence square integrable. Its mean is $R^n/\det\Lambda$, and its Fourier coefficient at $\xi\in\Lambda^*$ has absolute value
\[
 \frac{R^n}{\det\Lambda}
 \prod_i|\sinc(R\xi_i)|.
\]
Parseval's identity proves \eqref{eq:parseval}, including convergence of the nonnegative series.

The second moment can also be computed directly, by expanding $J_R^2$ before integrating over a fundamental domain. The two lattice points differ by some $w\in\Lambda$, and the intersection of the corresponding boxes has volume $\prod_i(R-|w_i|)_+$. Thus
\[
 \E[J_R^2]=
 \frac1{\det\Lambda}\sum_{w\in\Lambda}
       \prod_i(R-|w_i|)_+.
\]
Combining the two expressions for the second moment gives \eqref{eq:tent-poisson}.
\end{proof}

\subsection{From a one-dimensional weight to a cube bound}
The overlap identity in \cref{lem:parseval} reduces the variance
estimate to a weighted sum over the original lattice. To bound this
sum, we normalize the lattice to $L/\lambda$ and use the weight $f_r$
at $r=R/\lambda$, where, for $r>0$,
\begin{equation}\label{eq:tent-A}
 f_r(x)=(r-|x|)_+,
 \qquad A(r)=\sum_{k\in\Z}f_r(k).
\end{equation}

\begin{figure}[htbp]
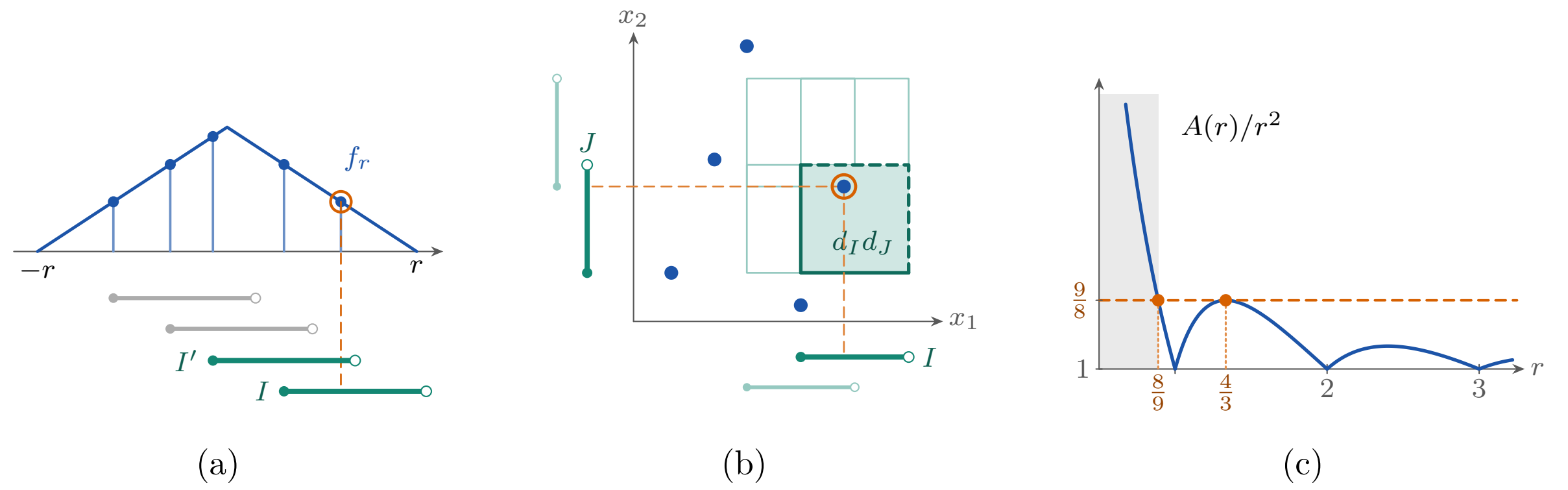

\centering
\begin{minipage}[b]{.31\textwidth}\centering
\FigPackCover\\[1pt]
{\small(a)}
\end{minipage}\hfill
\begin{minipage}[b]{.30\textwidth}\centering
\FigPackBoxes\\[1pt]
{\small(b)}
\end{minipage}\hfill
\begin{minipage}[b]{.35\textwidth}\centering
\FigPackRatio\\[1pt]
{\small(c)}
\end{minipage}
\caption{Weighted packing for cubes (\cref{lem:weighted-cube}).
(a) Weighted unit intervals cover the projected tent weights. (b) Each
product of cover intervals contains at most one point of $P$. (c) For
$r\ge8/9$, the ratio $A(r)/r^2$ attains its maximum $9/8$ at the two
marked points.}\label{fig:packing}
\end{figure}

The normalized lattice has separation at least one in the infinity
norm, so we seek a bound of $A(r)^n$ on the total product weight of any
set with this separation. Its coordinate projections need not
themselves be separated, so we first cover their weights by weighted
unit intervals. The next lemma relates the total cover weight to the
maximum weight of a separated subset, and the following lemma bounds
this maximum for $f_r$. Taking products of the covers gives half-open
unit cubes, each containing at most one point of the original set.
\Cref{fig:packing} illustrates the three steps.

\begin{lemma}\label{lem:interval-cover}
Let $x_1<\cdots<x_m$ be real numbers with nonnegative weights $w_1,\ldots,w_m$. There are nonnegative weights on the intervals $[x_j,x_j+1)$ that cover each point $x_i$ to total weight at least $w_i$. Their total weight equals the maximum weight of a subset whose points are pairwise at distance at least one.
\end{lemma}
\begin{proof}
Let $p(i)$ be the last index with $x_{p(i)}\le x_i-1$, or zero if no such index exists. Define
\[
 F_0=0,\qquad F_i=\max\{F_{i-1},w_i+F_{p(i)}\}.
\]
The first option omits $x_i$, while the second selects $x_i$ and uses an optimal subset among the first $p(i)$ points. Hence $F_m$ is the maximum weight of a separated subset.

Give $[x_j,x_j+1)$ weight $d_j=F_j-F_{j-1}\ge0$. The intervals containing $x_i$ are precisely those with $p(i)<j\le i$, so their total weight is
\[
 \sum_{j=p(i)+1}^i d_j=F_i-F_{p(i)}\ge w_i.
\]
The total interval weight is $\sum_jd_j=F_m$.
\end{proof}

\Needspace{10\baselineskip}
We now apply this covering principle to $f_r$ using the following
bound.

\begin{lemma}\label{lem:tent}
For every finite set $P\subset\R$ whose distinct points are at distance at least one,
\[
 \sum_{x\in P}f_r(x)\le A(r).
\]
Writing $r=m+u$ with $m=\lfloor r\rfloor$ and $0\le u<1$, one has
\[
 A(r)=r^2+u(1-u).
\]
\end{lemma}
\begin{proof}
Discard points with zero weight. If $k$ points remain, order them increasingly. Pair the smallest with the largest, the second smallest with the second largest, and so on. The two points of the $j$th pair are at distance at least $k+1-2j$, so
\[
 \sum_{x\in P}|x|\ge(k-1)+(k-3)+\cdots=\lfloor k^2/4\rfloor.
\]
Consequently their total weight is at most $kr-\lfloor k^2/4\rfloor$. For even $k$, this expression is the average of its values at the two neighboring odd integers. For $k=2j+1$, it equals $(2j+1)r-j(j+1)$. Maximizing over $j\ge0$ gives $A(r)$. Summing the terms at $0,\pm1,\ldots,\pm m$ gives the displayed formula.
\end{proof}

\begin{lemma}[Weighted packing for cubes]\label{lem:weighted-cube}
If $P\subset\R^n$ is $\ell_\infty$-separated by at least one, then
\begin{equation}\label{eq:weighted-cube}
 \sum_{x\in P}\prod_{i=1}^nf_r(x_i)\le A(r)^n.
\end{equation}
In particular, with $\lambda=\lambda_1^{(\infty)}(L)$,
\begin{equation}\label{eq:cube-theta}
 1+\Theta_R(L^*)
 \le (9/8)^n\frac{\det L}{\lambda^n}
 \qquad\text{for every }R\ge8\lambda/9.
\end{equation}
\end{lemma}
\begin{proof}
Replacing $P$ by $P\cap(-r,r)^n$ does not change the sum, so we may
assume that $P$ is finite and contained in this box. In each coordinate
$i$, apply \cref{lem:interval-cover} to the distinct values of the
$i$th coordinates of the points of $P$, with the weights given by $f_r$,
as in \cref{fig:packing}(a). This gives half-open unit intervals $I$ with
weights $d^{(i)}_I\ge0$ such that
\[
 f_r(x_i)\le\sum_{I\ni x_i}d^{(i)}_I\quad(x\in P),
 \qquad
 \sum_I d^{(i)}_I\le A(r).
\]
The second bound holds because the cover's total weight equals that of
the heaviest separated subset of the projection, which is at most
$A(r)$ by \cref{lem:tent}.

Multiply the coordinate covers. A product $K=I_1\times\cdots\times I_n$
of cover intervals is a half-open unit cube, to which we give the
weight $d_K=\prod_i d^{(i)}_{I_i}$. Two points of $K$ differ by less than
one in every coordinate, so $K$ contains at most one point of $P$, as
in \cref{fig:packing}(b). Expanding the product of the coordinate
bounds and exchanging the order of summation therefore gives
\[
 \sum_{x\in P}\prod_{i=1}^n f_r(x_i)
 \le\sum_{x\in P}\sum_{K\ni x}d_K
 =\sum_K d_K\,\#(P\cap K)
 \le\sum_K d_K
 =\prod_{i=1}^n\sum_I d^{(i)}_I
 \le A(r)^n,
\]
proving \eqref{eq:weighted-cube}.

To prove \eqref{eq:cube-theta}, set $r=R/\lambda$ and apply \eqref{eq:weighted-cube} to $L/\lambda$. Equation \eqref{eq:tent-poisson} gives
\[
 1+\Theta_R(L^*)\le
 \frac{\det L}{\lambda^n}\left(\frac{A(r)}{r^2}\right)^n.
\]
For $8/9\le r\le1$, the ratio is $1/r\le9/8$, which is where the threshold $8/9$ comes from. For $r=m+u\ge1$, it is at most
\[
 1+\frac{u(1-u)}{(1+u)^2}\le\frac98,
\]
since $8u(1-u)\le(1+u)^2$ is equivalent to $(3u-1)^2\ge0$.
\end{proof}

For $L=\Z^n$ and $R=4/3$, equality holds in \eqref{eq:cube-theta}, the second point of equality in \cref{fig:packing}(c). This example explains why the factor $9/8$ enters the list size.

\subsection{Exact moments for finite shifts}
A sufficiently aligned rational grid preserves the same moments exactly when shifts are sampled from the finite quotient used by the algorithm.

\begin{lemma}\label{lem:grid-moments}
Suppose $\Gamma\subseteq G=\delta\Z^n$ and $R/\delta\in\Z$. Fix $z\in G$. For uniform $a\in G/\Gamma$, let
\[
 J_a(z)=\#\bigl((a-z/2+\Gamma)\cap B_R\bigr),
 \qquad \mu_R=R^n/D.
\]
Then
\begin{align}
 \E_a[J_a(z)]&=\mu_R,\label{eq:grid-mean}\\
 \E_a[J_a(z)^2]&=(1+\Theta_R(\Gamma^*))\mu_R^2,\label{eq:grid-second}\\
 \E_a[J_a(z)J_a(z')]&\le
 (1+\Theta_R(\Gamma^*))\mu_R^2
 \qquad(z,z'\in G).\label{eq:grid-mixed}
\end{align}
\end{lemma}
\begin{proof}
Every interval of length $R$ on the translate $\delta\Z-z_i/2$ contains exactly $R/\delta$ points. Averaging the coset therefore gives
\[
 \E_a[J_a(z)]=\frac{\delta^n}{D}(R/\delta)^n=\mu_R.
\]
In the square of the count, the difference between two midpoints belongs to $\Gamma$. For $w\in\Gamma$, each coordinate intersection contains exactly $(R-|w_i|)_+/\delta$ points. Its endpoints differ by integer multiples of $\delta$, even when $z_i/2\notin\delta\Z$. Thus
\[
 \E_a[J_a(z)^2]=\frac1D\sum_{w\in\Gamma}
                \prod_i(R-|w_i|)_+.
\]
Use \eqref{eq:tent-poisson} for the second identity. The two second moments are the same for all $z$, so Cauchy--Schwarz proves the mixed bound.
\end{proof}

\section{A random superlattice with small box variance}\label{sec:super}
To use the moment identities from \cref{sec:cubes} in the list
recursion, we need a superlattice whose relative box variance is small
at every required width. The weighted cube estimate determines the
index needed to achieve this. Fix a candidate scale $\ell$ satisfying
\begin{equation}\label{eq:correct-scale}
 \ell\le\lambda\le(1+\eta)\ell,
 \qquad \eta=n^{-3}.
\end{equation}
Among the polynomially many scales tried by the algorithm, we analyze
one satisfying \eqref{eq:correct-scale}.

We use $h\ge2$ to control the variance tolerance and the failure
probability, and will choose $h=O(\log n)$ in \cref{sec:algorithm} to
accommodate the full recursion. Put
\begin{equation}\label{eq:prime-scale}
 C=\frac98,\qquad
 P_0=C^n\frac{\det L}{\ell^n}2^{4h}.
\end{equation}
The first factor reflects the weighted cube estimate, while the
additional factor $2^{4h}$ provides a margin for controlling all
required widths simultaneously.
Choose a prime $p$ with $P_0\le p\le4P_0$. If $L=B\Z^n$, choose a uniform nonzero $u\in\F_p^n$ and define
\begin{equation}\label{eq:superlattice}
 \Gamma=L+\Z\frac{Bu}{p},\qquad D=\det\Gamma=\frac{\det L}{p}.
\end{equation}
Then
\begin{equation}\label{eq:D-bounds}
 \frac14\ell^nC^{-n}2^{-4h}
 \le D\le\ell^nC^{-n}2^{-4h}.
\end{equation}
The index is always $p$, so $D$ is fixed before $u$ is drawn.

\subsection{Reducing box variance through the dual lattice}
The dual of \eqref{eq:superlattice} is
\[
 \Gamma^*=\{B^{-T}z:z\in\Z^n,\ \langle z,u\rangle=0\pmod p\}.
\]
A vector outside $pL^*$ survives with probability
\[
 \alpha_p=\frac{p^{n-1}-1}{p^n-1}<\frac1p,
\]
while every vector in $pL^*$ survives. Consequently,
\begin{equation}\label{eq:random-theta}
 \E_\Gamma[\Theta_R(\Gamma^*)]
 =\alpha_p\Theta_R(L^*)+(1-\alpha_p)\Theta_R(pL^*)
 \le\frac{\Theta_R(L^*)}{p}+\Theta_R(pL^*).
\end{equation}
This calculation was used for Gaussian weights in~\cite{GFH26a}, and
here we apply it to products of squared sinc functions.

The first term of \eqref{eq:random-theta} is at most $2^{-4h}$ by \cref{lem:weighted-cube}. We bound the second term, the variance associated with the fine lattice $L/p$, by counting fundamental cells near the boundary of a box.

\begin{lemma}\label{lem:forced}
For all sufficiently large $n$, the parameters above satisfy
\[
 \Theta_R(pL^*)\le2^{-4h}\qquad(R\ge\ell).
\]
The threshold is uniform over the input lattice, the correct scale, and $h\ge2$.
\end{lemma}
\begin{proof}
By \eqref{eq:minkowski},
\[
 \lambda_n^{(2)}(L)
 \le\sqrt n\frac{\det L}{\lambda^{n-1}}.
\]
By \cref{lem:basis} and \eqref{eq:cell-radius}, $L$ has a basis whose fundamental cell has coordinate radius at most $n^2\Ared(n)\lambda_n^{(2)}(L)$. Dividing this basis by $p$ gives a fundamental cell for $L/p$ contained in $[-F,F]^n$, where, with $K(n)=n^3\Ared(n)$,
\begin{equation}\label{eq:F-forced}
 F\le K(n)\ell C^{-n}2^{-4h}.
\end{equation}

Let $d=\det(L/p)$. For every translate of a box of side $R$, the union of the cells anchored at its lattice points is contained in the box expanded by $F$ in each direction. Conversely, it covers the box contracted by $F$ in each direction. Therefore
\[
 (R-2F)_+^n\le dJ_R(a)\le(R+2F)^n.
\]
For $2nF/R\le1$, the relative deviation from the mean $R^n/d$ is at most $4nF/R$. The variance identity in \cref{lem:parseval} gives
\[
 \Theta_R(pL^*)\le16n^2(F/R)^2
 \le16n^2K(n)^2C^{-2n}2^{-8h}.
\]
The last expression is at most $2^{-4h}$ for all sufficiently large $n$.
\end{proof}

\begin{theorem}[Small box variance on a random superlattice]\label{thm:super}
Fix $m$ prescribed widths $R\ge\ell$. For all sufficiently large $n$, the construction \eqref{eq:superlattice} satisfies
\[
 \Theta_R(\Gamma^*)\le\theta:=2^{-h}
 \quad\text{at every prescribed width}
\]
except with probability at most $m2^{1-3h}$.
\end{theorem}
\begin{proof}
Equation~\eqref{eq:random-theta}, together with
\cref{lem:weighted-cube,lem:forced}, gives
\[
 \E_\Gamma[\Theta_R(\Gamma^*)]\le2^{1-4h}.
\]
Apply Markov's inequality at each prescribed width and take a union bound.
\end{proof}

\subsection{Choosing the initial scale}
To initialize the recursion, we will sample directly in a cube much
wider than a fundamental cell of $\Gamma$, so we need a basis whose
cell radius is at most a polynomial factor times $\ell$. The small
variance bound provides such a basis through a lower bound on the
shortest dual vector and transference.

Call the event in \cref{thm:super} $\G$. We include $\ell$ among the prescribed widths. On $\G$, there is no nonzero $\xi\in\Gamma^*$ with $\ell\eunorm\xi\le1/4$. Indeed, the elementary inequality $\sin t\ge t-t^3/6$ gives
\[
 \prod_i\sinc^2(\ell\xi_i)
 \ge1-\frac{\pi^2}{3}\ell^2\eunorm\xi^2
 \ge1-\frac{\pi^2}{48}>\frac12,
\]
contradicting $\Theta_\ell(\Gamma^*)\le2^{-h}$. By \eqref{eq:transference},
\[
 \lambda_n^{(2)}(\Gamma)<4n\ell.
\]
After \cref{lem:basis}, the fundamental cell has coordinate radius at most
\begin{equation}\label{eq:Fmax}
 F_{\max}=4n^3\Ared(n)\ell.
\end{equation}
This bound can be checked from the computed basis by summing the infinity norms of its columns.

The root will have radius $R_0$, and the leaves will have the largest
radius $R_t$. Fix
\begin{equation}\label{eq:schedule}
 \gamma=\frac{19}{20},\qquad R_0=(1+\eta)\ell,
 \qquad R_j=R_0\gamma^{-j}.
\end{equation}
Let $t$ be the least nonnegative integer such that
\begin{equation}\label{eq:depth}
 R_t\ge8n^5\Ared(n)\ell=2n^2F_{\max}.
\end{equation}
Since $\Ared$ is a fixed polynomial, $t=O(\log n)$. Both $t$ and the ratios $R_j/\ell$ depend only on $n$. The widths supplied to \cref{thm:super} are $\ell,R_0,\ldots,R_t$, so they are fixed before the random superlattice is drawn.

\section{Building the lists}\label{sec:lists}
Fix $\Gamma\in\G$ and its preprocessed basis. We construct lists
satisfying the reference counting bounds in \eqref{eq:reference} along
a binary tree with leaves at radius $R_t$. Direct sampling establishes
the bounds at the leaves, while each shortening step preserves the
exact first moment and propagates the bound on the second counting
moment. Combining these steps gives a list in $(c+\Gamma)\cap Q_{R_j}$
for any prescribed coset. Each subtree uses private samples, private
shifts, and private retention coins.

To align the midpoint boxes and the coordinate partitions used by the
searches, choose a single grid $G=\delta\Z^n$ satisfying
\begin{equation}\label{eq:aligned-grid}
 p^{-1}L\subseteq G,\qquad
 \ell/(2\delta)\in\Z,\qquad
 R_j/(6\delta)\in\Z\quad(0\le j\le t),\qquad
 \delta/\ell\le n^{-10}.
\end{equation}
A common denominator followed by integer subdivision gives such a grid with polynomial encoding length.

\subsection{The leaf sampler}
At a leaf, the box is wide relative to a fundamental cell, allowing direct sampling with exactly constant expected counts.

Let $P=B_\Gamma[0,1)^n$ be the fundamental cell and use the bound $P\subseteq[-F_{\max},F_{\max}]^n$. Set
\[
 U=Q_{R_t+F_{\max}},\qquad V_U=\vol(U).
\]
Draw $X$ uniformly from $U\cap G$. For the prescribed coset $c+\Gamma$, compute
\begin{equation}\label{eq:rounding}
 Y=c+B_\Gamma\left\lfloor B_\Gamma^{-1}(X-c)\right\rfloor.
\end{equation}
Then $X\in Y+P$. Keep the proposal only when $Y\in Q_{R_t}$.

\begin{lemma}\label{lem:leaf}
For any positive integer $N$, let
\[
 M=\left\lceil N\frac{V_U}{(2R_t)^n}\right\rceil,
 \qquad q_0=\frac{NV_U}{M(2R_t)^n}.
\]
Make $M$ independent proposals as in \eqref{eq:rounding}. Independently keep each successful proposal with probability $q_0$. For every fixed coset $c+\Gamma$, the resulting list satisfies
\[
 m(x)=A_t:=\frac{ND}{(2R_t)^n},\qquad
 S(x,x')\le A_t^2
 \quad(x,x'\in(c+\Gamma)\cap Q_{R_t}).
\]
It uses $O(N)$ proposals, each of polynomial cost.
\end{lemma}
\begin{proof}
For every allowed output $y$, its complete cell $y+P$ lies in $U$. That cell contains exactly $D/\delta^n$ grid points, one for each coset of $\Gamma$ in $G$. Hence a proposal equals $y$ with probability $D/V_U$, independently of $y$ and $c$.

The first moment is $Mq_0D/V_U=A_t$. Distinct retained entries come from distinct proposals, so
\[
 S(x,x')=M(M-1)(q_0D/V_U)^2
 =(1-1/M)A_t^2.
\]
Finally, \eqref{eq:depth} gives
\[
 \frac{V_U}{(2R_t)^n}
 =\left(1+\frac{F_{\max}}{R_t}\right)^n
 \le e^{1/(2n)}.
\]
Thus $M=O(N)$.
\end{proof}

\subsection{One shortening step}
Suppose a node is to produce a list in $c+\Gamma$ at radius $\gamma R$. Choose a uniform $a\in G/\Gamma$. The two child cosets are
\begin{equation}\label{eq:child-cosets}
 c_+=a+\Gamma,\qquad c_-=a-c+\Gamma.
\end{equation}
Generate the two child lists at radius $R$ with independent private randomness. For each ordered pair of child entries, keep the difference with probability
\begin{equation}\label{eq:merge-kernel}
 q\ind_{\{x-y\in Q_{\gamma R}\}}
   \ind_{\{(x+y)/2\in B_R\}},
 \qquad q=\frac{(2/\gamma)^n}{N}.
\end{equation}
Each distinct pair of child identifiers uses an independent coin and creates at most one output entry.

Fix $z\in(c+\Gamma)\cap Q_{\gamma R}$. An admissible pair with difference $z$ is uniquely determined by its midpoint. Its midpoint set is
\[
 (a-z/2+\Gamma)\cap B_R.
\]
Every such midpoint gives two endpoints in the child boxes, so the
number of representations is exactly the variable $J_a(z)$ of
\cref{lem:grid-moments}.

The counting recursion of~\cite[Lemma~4.1]{GDD} applies here with a
constant expected count at each lattice point of the cube. The
additional error in the second counting moment comes from variation
in the midpoint count and reuse of an input entry. The box variance
controls the former, while the latter contributes a term that
decreases with the list size.

\begin{lemma}[Counting moments after a difference step]\label{lem:merge}
Suppose both child lists satisfy \eqref{eq:reference} with common expected count $A=A_R$ and error $\varepsilon$, uniformly over their prescribed cosets. Assume they are independent after the cosets in \eqref{eq:child-cosets} are fixed, and that $q\le1$. The output satisfies
\[
 m_{\mathrm{out}}(z)=A_{\gamma R}
\]
and
\begin{equation}\label{eq:merge-error}
 S_{\mathrm{out}}(z,z')\le e^{\varepsilon'}A_{\gamma R}^2,
 \qquad
 \varepsilon'=2\varepsilon+
 \log\left(1+\Theta_R(\Gamma^*)+\frac{2^{n+1}}N\right).
\end{equation}
The bounds hold for every fixed output coset and all
$z,z'\in(c+\Gamma)\cap Q_{\gamma R}$.
\end{lemma}
\begin{proof}
Fix $a$. The first moment at $z$ is $qA^2J_a(z)$. Averaging $a$ gives
\[
 qA^2\mu_R
 =\frac{(2/\gamma)^n}{N}
   \left(\frac{ND}{(2R)^n}\right)^2\frac{R^n}{D}
 =\frac{ND}{(2\gamma R)^n}=A_{\gamma R}.
\]

For the second factorial moment, classify two distinct output entries by their child identifiers. There are three cases: four distinct child entries, one shared left entry, or one shared right entry.

Four distinct child entries contribute at most
\[
 e^{2\varepsilon}q^2A^4\E_a[J_a(z)J_a(z')]
 \le e^{2\varepsilon}q^2A^4(1+\Theta_R)\mu_R^2.
\]
For a shared left entry at $x$, the right values are fixed as $x-z$ and $x-z'$. Since the number of possible triples is at most $J_a(z)$, this contribution is at most
\[
 e^{2\varepsilon}q^2A^3\E_a[J_a(z)]
 =e^{2\varepsilon}q^2A^3\mu_R.
\]
The same bound holds for a shared right entry. This also covers $z=z'$, since different entries may have equal values. Sharing both child identifiers would give the same output entry and is excluded from the factorial moment.

Divide the sum by $A_{\gamma R}^2=(qA^2\mu_R)^2$. Since
\[
 A\mu_R=N/2^n,
\]
the result is \eqref{eq:merge-error}.
\end{proof}

\subsection{The complete tree}
Set $A_j=ND/(2R_j)^n$. A call $\Build(j,c)$ produces a list in $(c+\Gamma)\cap Q_{R_j}$. At $j=t$, it invokes the leaf sampler. Otherwise, it uses \eqref{eq:child-cosets}, calls $\Build(j+1,a)$ and $\Build(j+1,a-c)$ independently, and applies \eqref{eq:merge-kernel} at input radius $R_{j+1}$.

Starting with $\varepsilon_t=0$, \cref{lem:merge} gives
\begin{equation}\label{eq:tree-error}
 \varepsilon_0\le(2^t-1)
 \left(\theta+\frac{2^{n+1}}N\right).
\end{equation}
The factor $2^t$ counts the effect of the entire tree. It remains polynomial because $t=O(\log n)$.

The first moment stays exact throughout the construction. If a node's coset is uniform in $G/\Gamma$, then \eqref{eq:one-membership} gives
\begin{equation}\label{eq:expected-size}
 \E[|\List_j|]
 =A_j\frac{\delta^n}{D}\#(Q_{R_j}\cap G)=N.
\end{equation}
A uniform root coset makes every node coset uniform.

\begin{algorithm}[htbp]
\caption{$\Build(j,c)$}\label{alg:build}
\KwIn{Level $j$, coset representative $c\in G$, fixed $\Gamma$, grid and radii.}
\KwOut{A list in $(c+\Gamma)\cap Q_{R_j}$, or $\Fail$.}
\If{$j=t$}{\Return the list of \cref{lem:leaf}}
Choose $a$ uniformly in $G/\Gamma$\;
$\List_+\leftarrow\Build(j+1,a)$ and $\List_-\leftarrow\Build(j+1,a-c)$, using independent randomness\;
\If{either call failed}{\Return $\Fail$}
Initialize an empty output list\;
Use the ternary search of \cref{sec:internal} on $\List_+,\List_-$\;
\ForEach{reported pair of entries $(x,y)$}{
 \If{$x-y\in Q_{R_j}$ and $(x+y)/2\in B_{R_{j+1}}$}{
  With probability $q$, append a new entry with value $x-y$\;
 }
}
\Return the output list\;
\end{algorithm}

\Cref{alg:build} describes the generation rule. The implementation places fixed limits on list sizes and search work. These limits are chosen in \cref{sec:algorithm}. Until then, the counting analysis refers to the same finite construction without those limits.

\FloatBarrier
\section{Finding pairs for recursive differencing}\label{sec:internal}
To implement the difference rule, we first select candidate pairs by
their coordinate classes and then test the exact conditions in
\eqref{eq:merge-kernel}. A prefix table reports these candidates, and
the counting identities from \cref{sec:lists} bound their expected
number.

\subsection{Three classes per coordinate}
At input radius $R$, define
\begin{equation}\label{eq:ternary-classes}
 b_i(x)=
 \begin{cases}
 0,& |x_i|\le R/3,\\
 1,& R/3<|x_i|\le2R/3,\\
 2,& 2R/3<|x_i|\le R.
 \end{cases}
\end{equation}
For an eligible pair,
\[
 |x_i|+|y_i|=\max\{|x_i-y_i|,|x_i+y_i|\}\le R,
\]
so $b_i(x)+b_i(y)\le2$. Thus, for each left entry $x$, we need to
report all right entries whose class vectors satisfy
$b_i(y)\le2-b_i(x)$ in every coordinate. We next give a table that
answers these queries and will also be used for the binary classes
in the final search.

\subsection{Prefix tables and search cost}\label{sec:prefix-tables}
Consider records with class vectors in $\{0,\ldots,s-1\}^n$, where
$s$ is either two or three. Given upper bounds
$u\in\{0,\ldots,s-1\}^n$, a query reports all records whose class
vector $b$ satisfies $b_i\le u_i$ in every coordinate.

Let $H(b)$ be the histogram of class vectors, with a list of record identifiers at each occupied vector. For every $0\le k\le n$, store a table
\begin{equation}\label{eq:prefix-table}
 T_k(a_1,\ldots,a_k,u_{k+1},\ldots,u_n)
 =\sum_{\substack{b_{k+1}\le u_{k+1},\ldots,b_n\le u_n}}
 H(a_1,\ldots,a_k,b_{k+1},\ldots,b_n).
\end{equation}
Each table has $s^n$ entries and is obtained from $H$ by prefix sums in the last $n-k$ coordinates. All tables therefore take $O(n^2s^n)$ arithmetic operations and $O(ns^n)$ storage, in addition to the records.

To answer a query, start with the empty prefix. At depth $k$, try the at most $s$ allowed values for the next coordinate and consult $T_{k+1}$. Descend only when the count is positive. At depth $n$, report the identifiers stored at that class vector.

Every visited nonempty prefix has a reported record below it. Charging
the prefix to the first such record gives at most $n$ charges per
record, each with at most $s$ child tests, so a query costs
\[
 O\bigl(ns(1+\text{number of reported records})\bigr).
\]

For the internal search, build the ternary tables on the right list
and query with $u_i=2-b_i(x)$ for each left entry $x$. Test the exact
difference and midpoint conditions on every report. Each pair of entry
identifiers is reported at most once, so independent retention coins
implement the generation rule of \cref{sec:lists}.

\begin{proposition}\label{prop:internal-work}
Fix $\Gamma\in\G$ and an internal call whose prescribed output coset is uniform in $G/\Gamma$. If the two child lists have the exact first moments in \eqref{eq:reference}, then their expected number of ternary candidate pairs is
\begin{equation}\label{eq:ternary-count}
 \E[P_3]=N^2\left(
 \frac23+\frac{2\delta}{3R}-\frac{\delta^2}{4R^2}
 \right)^n.
\end{equation}
Including construction of the tables, the expected search work is at most
\[
 \poly(n,\bits(B))\bigl(3^n+N+N^2(2/3)^n e^{O(n\delta/R)}\bigr).
\]
\end{proposition}
\begin{proof}
Write the parent coset as $c+\Gamma$. For independent uniform $c,a\in G/\Gamma$, the map
\[
 (c,a)\longmapsto(a,a-c)
\]
is a bijection of the quotient squared. Hence the two child cosets are jointly uniform. First fix these cosets and multiply the child first moments, using their private independence. Averaging the cosets then shows that every pair in $(Q_R\cap G)^2$ has expected count
\[
 A_R^2(\delta^n/D)^2.
\]

Let $m=R/\delta$, which is divisible by six. The three one-dimensional class sizes are $2m/3+1$, $2m/3$, and $2m/3-1$. Summing the six combinations with class sum at most two gives the fraction
\[
 \frac23+\frac{2}{3m}-\frac1{4m^2}.
\]
The free grid has $(2R/\delta)^{2n}$ endpoint pairs, which proves \eqref{eq:ternary-count} after multiplication by the expected count above. The table construction and query bounds, together with $\E[|\List_\pm|]=N$, give the work estimate.
\end{proof}

A uniform common root coset makes the output coset of every internal call marginally uniform, so \cref{prop:internal-work} applies at every level.

\section{Recovering a shortest vector}\label{sec:terminal}
The two final lists lie in the same coset $z+\Gamma$ and are independent
conditional on $z$ and $\Gamma$. The objective now changes from
generating another list to recovering a shortest vector of $L$ from
their differences. We combine quotient labels that identify
differences in $L$ with a binary condition that uses the separation of
$L$ to control the expected number of reports. After
defining the condition, we show that it retains enough representations
of a shortest vector, bound all reported pairs, and give the search
algorithm.

\subsection{Quotient labels and a binary condition}
For $x=z+g_x$ with $g_x\in\Gamma$, give the entry label
\begin{equation}\label{eq:label}
 \lab(x)=g_x+L\in\Gamma/L.
\end{equation}
Then
\[
 \lab(x)=\lab(y)\quad\Longleftrightarrow\quad x-y\in L.
\]
A label is represented by one field element, as detailed in Appendix~\ref{app:implementation}.

Let
\[
 I=[-R_0,R_0),\qquad B_0=[-\ell/2,\ell/2).
\]
For each endpoint, store the binary class vector
\[
 h_i(x)=\ind_{\{x_i\notin B_0\}}.
\]
The final search reports exactly the pairs of entries satisfying
\begin{equation}\label{eq:binary-condition}
 \lab(x)=\lab(y),\qquad
 h_i(x)+h_i(y)\le1\quad(1\le i\le n).
\end{equation}
Thus every coordinate has at least one endpoint in the inner interval $B_0$. Each reported nonzero difference is tested for membership in $L$ and compared with the current best vector.

\subsection{A shortest vector has many compatible representations}
Fix a shortest vector $v$. The full midpoint box $B_\ell$ gives endpoints
\[
 x=M+v/2,\qquad y=M-v/2
\]
in $Q_{R_0}$. Let $J_{\mathrm{full}}(z)$ count its midpoints in $z-v/2+\Gamma$, and let $J(z)$ count those also satisfying the binary condition in \eqref{eq:binary-condition}.

\begin{lemma}\label{lem:target-midpoints}
For every $\Gamma\in\G$, a uniform $z\in G/\Gamma$ satisfies
\begin{equation}\label{eq:target-J}
 J(z)\ge\frac14\frac{\ell^n}{D}
\end{equation}
except with probability at most $4\theta+4n\eta$.
\end{lemma}
\begin{proof}
Put $\mu=\ell^n/D$. By \cref{lem:grid-moments},
\[
 \E_z[J_{\mathrm{full}}]=\mu,\qquad
 \Var_zJ_{\mathrm{full}}\le\theta\mu^2.
\]
Chebyshev's inequality gives $J_{\mathrm{full}}\ge\mu/2$ except with probability $4\theta$.

Consider one coordinate. If $|v_i|\le\ell$, every midpoint in $[-\ell/2,\ell/2)$ puts at least one endpoint in $B_0$. If $|v_i|>\ell$, the excluded midpoints form a central interval of length $|v_i|-\ell\le\eta\ell$. The endpoints of this interval align with the midpoint grid because $v_i\in\delta\Z$ and $\ell/(2\delta)\in\Z$.

A union bound over coordinates gives an expected deletion count of at most $n\eta\mu$, so Markov's inequality bounds the probability of deleting more than $\mu/4$ by $4n\eta$. On the complementary events, \eqref{eq:target-J} holds.
\end{proof}

To turn these midpoint locations into actual pairs, fix a root coset
with $J\ge1$ and let $Z$ count pairs of entries with difference $v$
and a compatible midpoint. The two root lists are independent with
the moments in \eqref{eq:reference}, so $\E[Z]=A_0^2J=:\mu_Z$.
Since each endpoint determines the other uniquely, two different
midpoint locations have different endpoint values on both sides and
contribute at most $e^{2\varepsilon_0}A_0^4$ to the second moment.
At a repeated location, the ordinary second moment on each side is
at most $e^{\varepsilon_0}(A_0^2+A_0)$. Consequently,
\[
 \E[Z^2]\le e^{2\varepsilon_0}
 \bigl(A_0^4J^2+2A_0^3J+A_0^2J\bigr),
\]
and the second-moment inequality gives
\begin{equation}\label{eq:target-prob}
 \Prb[Z>0]\ge
 \frac{e^{-2\varepsilon_0}}{1+2/(A_0J)+1/\mu_Z}
 \ge\frac{e^{-2\varepsilon_0}}{(1+\mu_Z^{-1/2})^2},
\end{equation}
where the last inequality uses $A_0J=\sqrt{\mu_ZJ}\ge\sqrt{\mu_Z}$.

On \eqref{eq:target-J}, the expected number of these pairs satisfies
\begin{equation}\label{eq:target-mass}
 A_0^2J\ge
 \frac{N^2D\ell^n}{4(2R_0)^{2n}}.
\end{equation}
The polynomial factors selected in \cref{sec:algorithm} make this quantity tend to infinity.

\subsection{Bounding the number of reported pairs}
\begin{figure}[htbp]
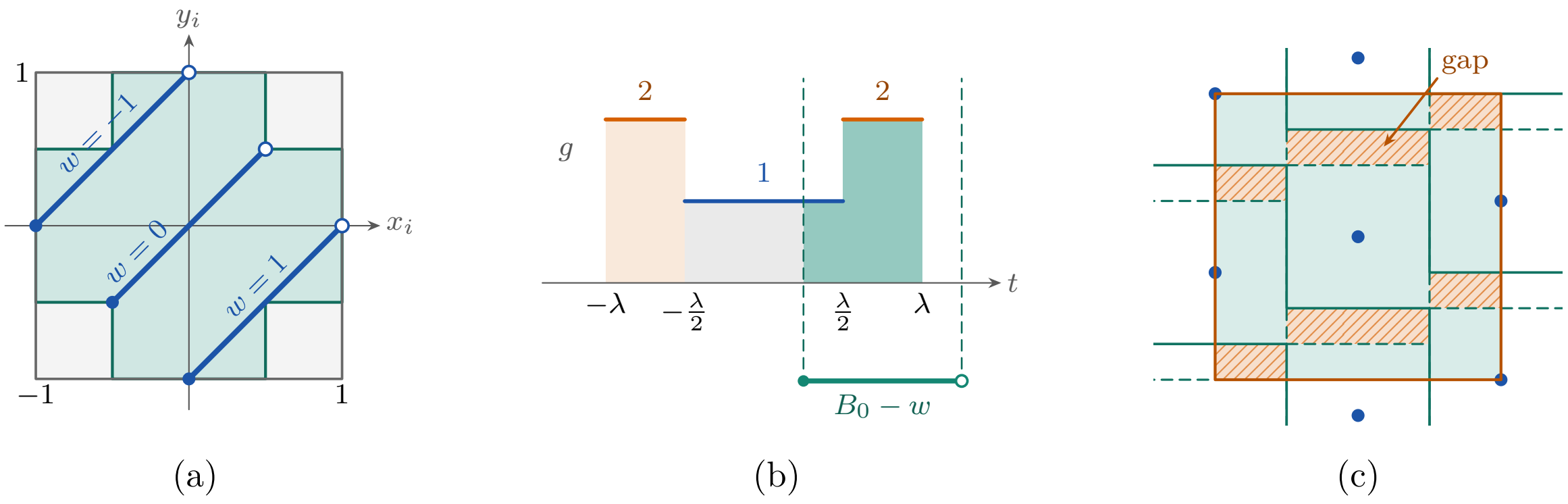

\centering
\begin{minipage}[b]{.31\textwidth}\centering
\FigWinSections\\[1pt]
{\small(a)}
\end{minipage}\hfill
\begin{minipage}[b]{.35\textwidth}\centering
\FigWinWeight\\[1pt]
{\small(b)}
\end{minipage}\hfill
\begin{minipage}[b]{.31\textwidth}\centering
\FigWinDisjoint\\[1pt]
{\small(c)}
\end{minipage}
\caption{Counting reported pairs at $R_0=\ell=\lambda$
(\cref{lem:terminal-reports}). (a) For $L=\Z$, each marked slice has
horizontal width one. (b) The mass of $g$ inside $B_0-w$ gives the
contribution at displacement $w$. (c) The windows $B_\ell-w$ with
$w\in L$ are disjoint. Hatching marks the gaps.}\label{fig:windows}
\end{figure}

We now bound all pairs satisfying \eqref{eq:binary-condition}, including
zero differences. The two lists lie in a common translate of $\Gamma$,
so every difference $w=x-y$ lies in $\Gamma$, a lattice with many points
in a cube of side $\ell$. Equal labels restrict $w$ to $L$, and it is
the separation of $L$ that bounds the count. We will write the number of compatible pairs at displacement
$w$ as the part of a fixed nonnegative weight that lies in the window
$B_\ell-w$, as illustrated in \cref{fig:windows}. Each translate of
$B_\ell$ contains at most one point of $L$, so the windows of different
$w\in L$ are disjoint and together collect at most the total weight. We
first derive the one-dimensional identity behind this expression.

Write
\[
 I_\delta=I\cap\delta\Z,\qquad
 B_\delta=B_0\cap\delta\Z.
\]
For $w\in\delta\Z$, let $c_\delta(w)$ be $\delta$ times the number of
compatible one-dimensional endpoint pairs at displacement $w$:
\begin{equation}\label{eq:c-delta}
 c_\delta(w)=\delta\sum_{x\in\delta\Z}
 \ind_{I_\delta}(x)\ind_{I_\delta}(x-w)
 \ind_{\{x\in B_\delta\ \mathrm{or}\ x-w\in B_\delta\}}.
\end{equation}
Inclusion and exclusion counts pairs with the first endpoint in
$B_\delta$, adds those with the second endpoint there, and subtracts
those with both endpoints there. The first two counts agree because
$I_\delta$ and $B_\delta$ are symmetric about the same point
$-\delta/2$. Thus, with
\[
 g=2\ind_{I_\delta}-\ind_{B_\delta}\ge0,
\]
we obtain
\begin{equation}\label{eq:correlation}
 c_\delta(w)=\delta\sum_{x\in\delta\Z}
                   \ind_{B_\delta}(x)g(x-w).
\end{equation}
The weight $g$ is one on $B_\delta$ and two on
$I_\delta\setminus B_\delta$, so its total mass is
\[
 \delta\sum_{x\in\delta\Z}g(x)=4R_0-\ell.
\]
\Needspace{10\baselineskip}
Taking products over coordinates gives the factor in the following
bound.

\begin{lemma}[Final report bound]\label{lem:terminal-reports}
If $\ell\le\lambda_1^{(\infty)}(L)$, then
\begin{equation}\label{eq:terminal-packing}
 \sum_{w\in L}\prod_{i=1}^n c_\delta(w_i)
 \le(4R_0-\ell)^n.
\end{equation}
For the two root lists generated in a uniform common coset of a fixed $\Gamma\in\G$, the number $P$ of pairs satisfying \eqref{eq:binary-condition} obeys
\begin{equation}\label{eq:terminal-P}
 \E[P]\le\frac{A_0^2}{D}(4R_0-\ell)^n.
\end{equation}
\end{lemma}
\begin{proof}
Apply \eqref{eq:correlation} in every coordinate, then set $s=x-w$. All summands are nonnegative, so
\[
 \sum_{w\in L}\prod_i c_\delta(w_i)
 =\delta^n\sum_{s\in G}\prod_i g(s_i)
            \sum_{w\in L}\ind_{B_\delta^n}(s+w).
\]
The inner sum is at most one. Two points in the half-open cube $B_\delta^n$ differ by less than $\ell$ in every coordinate, while a nonzero vector of $L$ has infinity norm at least $\ell$. Therefore
\[
 \sum_{w\in L}\prod_i c_\delta(w_i)
 \le\delta^n\sum_{s\in G}\prod_i g(s_i)
 =\left(\delta\sum_{x\in\delta\Z}g(x)\right)^n
 =(4R_0-\ell)^n.
\]

For the work estimate, first fix the common coset $z+\Gamma$. The two root lists use independent private randomness, and their first moments are exactly $A_0$. If $x-y=w\in L$, then membership of $x$ in $z+\Gamma$ already implies membership of $y$. Averaging $z$ therefore contributes one factor $\delta^n/D$.

For a fixed $w$, the number of compatible free grid pairs is $\delta^{-n}\prod_i c_\delta(w_i)$. Thus
\[
 \E[P]=\frac{A_0^2}{D}
       \sum_{w\in L}\prod_i c_\delta(w_i),
\]
which proves \eqref{eq:terminal-P}.
\end{proof}

To see the $3^n$ balance already at this point, put $r=R_0/\ell$.
Our parameters give $N^2D/\ell^n=4^n\poly(n)$, so the report bound is
\[
 \E[P]\le\poly(n)\left(\frac{4r-1}{r^2}\right)^n
 \le3^n\poly(n),\qquad r\ge1.
\]
The last inequality is $4r-1\le3r^2$. The remaining task is to find
these pairs without scanning all pairs in a large label group.

\subsection{Searching within label groups}
To turn the report bound into a running time bound, we must also control the
cost of searching within each label group. A split by group size
balances direct comparisons against the cost of building a binary table.

\begin{proposition}\label{prop:labeled-search}
Let two fixed labeled lists contain $M$ entries in total, and let $P$ be the number of their pairs satisfying \eqref{eq:binary-condition}. There is a deterministic algorithm that reports every such pair of entries once, in time
\begin{equation}\label{eq:terminal-search-time}
 \poly(n,\log(M+2),\text{label length})
 \bigl(M2^{n/2}+P\bigr)
\end{equation}
and space $\poly(n)(M+2^n)$.
\end{proposition}
\begin{proof}
Sort the entries by label. Put $H=\lceil2^{n/2}\rceil$, and let $m_\xi$ be the total size of the two sides in label group $\xi$.

If $m_\xi\le H$, check all cross pairs in that group. The total cost of these groups is at most $\sum_\xi Hm_\xi\le HM$.

There are at most $M/H$ groups with $m_\xi>H$. In each such group, build the binary prefix tables of \cref{sec:prefix-tables} on the right side and query with bounds $u_i=1-h_i(x)$ for each left entry $x$. The total table and query cost is
\[
 \poly(n)\left(\frac M H2^n+M+P\right).
\]
Together with the small groups and sorting, this gives \eqref{eq:terminal-search-time}. Process the large groups one at a time to obtain the space bound.
\end{proof}

\section{The algorithm and its analysis}\label{sec:algorithm}
We now choose the polynomial factors, remove knowledge of $\lambda$, and place finite limits on every trial. The list construction and the two search bounds then give the claimed complexity.

\subsection{Parameters and the scale grid}
The constants and radii are those of \eqref{eq:schedule} and
\eqref{eq:depth}. We choose the variance tolerance and the polynomial
factor in the list size together to keep the accumulated moment error
small and make the expected number of target pairs large. Let
\begin{equation}\label{eq:polynomial-parameters}
 Q=2^{t+2},\qquad
 h=\left\lceil\log_2\bigl(Q(n+2)^{10}\bigr)\right\rceil,
 \qquad
 N=2^{3h}\left\lceil\sqrt{(9/2)^n}\right\rceil.
\end{equation}
Here $Q$ exceeds the total number of nodes in the two trees. Since $t=O(\log n)$,
\[
 Q=\poly(n),\qquad h=O(\log n),\qquad
 (3/\sqrt2)^n2^{3h}\le N\le2(3/\sqrt2)^n2^{3h}.
\]
Since $\gamma\sqrt C>1$, the retention probability in \eqref{eq:merge-kernel} satisfies
\[
 q\le2^{-3h}(\gamma\sqrt C)^{-n}<1.
\]
On every good superlattice, \eqref{eq:tree-error} gives
\begin{equation}\label{eq:eps-final}
 \varepsilon_0\le Q\left(2^{-h}+\frac{2^{n+1}}N\right)
 \le (n+2)^{-10}+2Q^{-2}(n+2)^{-30}.
\end{equation}

Set
\[
 \ell_{\mathrm{lo}}=\|B^{-1}\|_{\infty\to\infty}^{-1},
 \qquad
 \ell_{\mathrm{hi}}=\min_i\infnorm{b_i}.
\]
Then $\ell_{\mathrm{lo}}\le\lambda\le\ell_{\mathrm{hi}}$. We try every
scale $\ell_j=\ell_{\mathrm{lo}}(1+\eta)^j\le\ell_{\mathrm{hi}}$.
This gives
\[
 O\bigl(1+n^3\log(\ell_{\mathrm{hi}}/\ell_{\mathrm{lo}})\bigr)
 =\poly(n,\bits(B))
\]
scales of polynomial encoding length, at least one of which satisfies
\eqref{eq:correct-scale}.

At each scale, set $Y=\max\{2,\lceil P_0\rceil\}$ and test
$O((n+h+1)\log(2Y))$ independent uniform candidates in $[Y,2Y]$ by
deterministic primality testing~\cite{AKS04}. Standard prime density
bounds make the failure probability at most $2^{-n-10}$, and every
prime returned on a correct scale satisfies $P_0\le p\le4P_0$.

\subsection{A trial with resource limits}
Choose the list limit
\begin{equation}\label{eq:list-cap}
 \Lcap=N2^{2h}.
\end{equation}
Let $p_{\mathrm{op}}(n,\bits(B))$ bound the polynomial overhead of the
arithmetic and search operations, and set the work limit for each
pair search to
\begin{equation}\label{eq:work-cap}
 \Wcap=3^n2^{8h}p_{\mathrm{op}}(n,\bits(B)).
\end{equation}
Stop the current trial before exceeding either limit, including the
cost of table construction and pair reporting. Retain every lattice
vector already found as a candidate.

\begin{algorithm}[htbp]
\caption{Shortest vector search in the infinity norm}\label{alg:main}
\KwIn{A nonsingular rational matrix $B$, with $L=B\Z^n$.}
\KwOut{A nonzero lattice vector $\Best$.}
Initialize $\Best$ with a shortest column of $B$ in the infinity norm\;
Compute $t,Q,h,N$ from \eqref{eq:depth} and \eqref{eq:polynomial-parameters}\;
\ForEach{scale $\ell$ in the multiplicative grid}{
 Compute $R_0,\ldots,R_t$, $F_{\max}$, and $P_0$\;
 Run the bounded prime search. Skip this scale if it fails\;
 Choose uniform nonzero $u\in\F_p^n$ and form $\Gamma=L+\Z Bu/p$\;
 Apply \cref{lem:basis} to $\Gamma$\;
 \If{the sum of the infinity norms of the basis columns exceeds $F_{\max}$}{Skip this scale}
 Choose $G=\delta\Z^n$ satisfying \eqref{eq:aligned-grid}, and choose uniform $z\in G/\Gamma$\;
 Run two independent calls $\List_+\leftarrow\Build(0,z)$ and $\List_-\leftarrow\Build(0,z)$ with limits \eqref{eq:list-cap} and \eqref{eq:work-cap}\;
 \If{either call fails}{Skip this scale}
 Assign quotient labels and run the search of \cref{prop:labeled-search}, with work limit $\Wcap$\;
 \ForEach{reported pair $(x,y)$}{
  Set $w=x-y$\;
  \If{$w\ne0$, $B^{-1}w\in\Z^n$, and $\infnorm w<\infnorm{\Best}$}{Set $\Best\leftarrow w$}
 }
}
\Return $\Best$\;
\end{algorithm}

We analyze \cref{alg:main} with exact random choices.
Appendix~\ref{app:implementation} implements these choices within fixed
resource bounds, with total additional failure probability at most
$2^{-n-10}$.

\subsection{Success of the unrestricted construction}
Fix a shortest vector $v$ and a correct scale. Conditional on a successful prime search, \cref{thm:super} gives
\[
 \Prb[\G^c]\le(t+2)2^{1-3h}=o(1).
\]
On $\G$, the basis check passes. By \cref{lem:target-midpoints}, the root shift has $J\ge\ell^n/(4D)$ except with probability $4\theta+4n\eta=o(1)$.

For every such root shift, combine \eqref{eq:D-bounds}, \eqref{eq:target-mass}, and \eqref{eq:polynomial-parameters}:
\begin{equation}\label{eq:target-M-lower}
 A_0^2J\ge
 \frac{2^{2h}}{16(1+\eta)^{2n}}=:\mu_*.
\end{equation}
Since $\mu_*\to\infty$, the moment bounds in \eqref{eq:reference} and
\eqref{eq:eps-final}, together with independence conditional on the
common root coset, allow us to apply \eqref{eq:target-prob}. The two
trees therefore contain a pair with difference $v$ satisfying the
binary condition with probability at least
\[
 \frac{e^{-2\varepsilon_0}}{(1+\mu_*^{-1/2})^2}=1-o(1).
\]
The deterministic final search reports every such pair unless its resource limit is reached.

\subsection{Bounding the probability of interruption}
All expectations in this subsection fix a good $\Gamma$ and average the root shift and the private tree randomness. They refer to the construction without resource limits.

Every node has expected list size $N$ by \eqref{eq:expected-size}. A union bound and Markov's inequality give
\begin{equation}\label{eq:list-overflow}
 \Prb[\text{some list exceeds }\Lcap]\le Q2^{-2h}=o(1).
\end{equation}
After absorbing constants into $p_{\mathrm{op}}$,
\cref{prop:internal-work} and $\delta/R\le n^{-10}$ bound the expected
work of each internal search by
\[
 p_{\mathrm{op}}(n,\bits(B))\,3^n2^{6h}.
\]
For the final search, \eqref{eq:terminal-P} and the parameter choices
give
\[
 \E[P]\le4\cdot2^{2h}
 \left(\frac{3+4\eta}{(1+\eta)^2}\right)^n
 \le4\cdot3^n2^{2h}.
\]
Also,
\[
 \E[M]\,2^{n/2}=2N2^{n/2}\le4\cdot3^n2^{3h},
\]
so \cref{prop:labeled-search} gives the same expected work bound for
the final search. The work limit \eqref{eq:work-cap} leaves a factor
of $2^{2h}$, and Markov's inequality followed by a union bound over
the searches gives
\begin{equation}\label{eq:work-overflow}
 \Prb[\text{some search reaches its work limit before a list limit}]
 \le O(Q2^{-2h})=o(1).
\end{equation}
These bounds do not condition on the presence of the target pair.

To compare the implementation with the unrestricted construction, give both the same random choices until the first limit is reached. They produce identical records before that point. Equations \eqref{eq:list-overflow} and \eqref{eq:work-overflow} bound the probability of this first interruption. Combined with the preceding correctness argument, one correct scale trial succeeds with probability $1-o(1)$.

\subsection{Total time and space}
There are polynomially many scales and at most $Q=\poly(n)$ tree
nodes per scale. Since $h=O(\log n)$, leaf sampling and pair searches
take $\OO(3^n)$ time under the stated resource limits, as summarized in
\cref{tab:balance}. The
$2^{n+o(n)}\poly(\bits(B))$ basis preprocessing cost per scale and
the polynomial overhead of the remaining operations are absorbed
in this bound.

Evaluating the trees in depth-first order keeps at most $O(t)$
completed lists waiting, so the live lists use $O(tN2^{2h})$ records.
An internal search uses $O(n3^n)$ table entries, while the final search
needs only $O(n2^n)$ additional table entries because large label groups
are processed one at a time. Since each record and each table count
has polynomial encoding length, the total space is $\OO(3^n)$.

\begin{proof}[Proof of \cref{thm:main}]
On a correct scale, the preceding bounds give total failure probability
$o(1)$, including the finite randomness loss. Thus \cref{alg:main}
succeeds with probability at least $2/3$ for all sufficiently large $n$.

When the correct scale succeeds, its final search reports a shortest vector. Every update of $\Best$ is an exactly checked nonzero vector of $L$, so a later scale cannot replace it by an invalid shorter vector. The finitely many smaller dimensions can use the $\ell_\infty$-SVP algorithm of~\cite{DPV11}, with their costs included in the absolute constant.
\end{proof}

\subsection{Enumerating all shortest vectors}\label{sec:enumeration}
Write $\mathcal S_\infty(L)=\{v\in L\setminus\{0\}:\infnorm v=\lambda\}$, and let $\mathcal C\subseteq L\setminus\{0\}$ be the set of nonzero lattice differences $w=x-y$ examined in one execution of \cref{alg:main}. For any fixed $v\in\mathcal S_\infty(L)$, the analysis of \cref{thm:main} bounds the probability that $v\notin\mathcal C$, giving $\Prb[v\in\mathcal C]=1-o(1)$ uniformly over $L$ and $v$, and hence $\Prb[v\in\mathcal C]\ge2/3$ for all sufficiently large $n$.

Mapping $\mathcal S_\infty(L)\cup\{0\}$ to $L/3L$ by $x\mapsto x+3L$ is injective, since $x-y=3w$ with $w\in L\setminus\{0\}$ would imply $\infnorm w\le(\infnorm x+\infnorm y)/3\le2\lambda/3<\lambda$. Thus $|\mathcal S_\infty(L)|\le|L/3L|-1=3^n-1$~\cite[Equation~(1.3)]{Henk02}.

\begin{proof}[Proof of \cref{thm:all-shortest}]
Set $m=n+1$. Initialize $\rho=\infnorm b$ and $\mathcal V=\{b\}$ with a shortest column $b$ of $B$, then perform $m$ independent runs of \cref{alg:main}, reusing the list and table workspace across runs. To maintain all vectors of the current minimum norm $\rho$, for each nonzero $w=x-y\in L$ examined in a terminal loop, replace $(\rho,\mathcal V)$ by $(\infnorm w,\{w\})$ if $\infnorm w<\rho$, and insert $w$ into $\mathcal V$ if $\infnorm w=\rho$ and $w\notin\mathcal V$. Output $\mathcal V$ after $m$ runs.

The output $\mathcal V$ is always a nonempty list of distinct nonzero vectors of $L$, and $\mathcal V=\mathcal S_\infty(L)$ whenever $\mathcal S_\infty(L)\subseteq\bigcup_{r=1}^m\mathcal C_r$. Because $(\rho,\mathcal V)$ does not affect the random choices or searches in any run, the sets $\mathcal C_1,\ldots,\mathcal C_m$ are independent. For all sufficiently large $n$, a union bound over $\mathcal S_\infty(L)$ therefore gives
\[
 \Prb[\mathcal V\ne\mathcal S_\infty(L)]
 \le\sum_{v\in\mathcal S_\infty(L)}\prod_{r=1}^m\Prb[v\notin\mathcal C_r]
 \le(3^n-1)3^{-m}<\frac13.
\]
In the finitely many smaller dimensions, the deterministic enumeration algorithm of~\cite{DPV11} returns a candidate set containing $\mathcal S_\infty(L)$ in $\poly(\bits(B))$ time, from which exact comparisons in the infinity norm retain $\mathcal S_\infty(L)$. Under \eqref{eq:list-cap} and \eqref{eq:work-cap}, each run examines $\OO(3^n)$ candidates, and maintaining $\mathcal V$ in a balanced search tree keyed by $B^{-1}w$ keeps every execution within $\OO(3^n)$ time and space.
\end{proof}

\section{Conclusion and discussion}\label{sec:conclusion}
We gave a classical randomized algorithm for exact $\ell_\infty$-SVP in
$\OO(3^n)$ time and space. Restricting midpoints to a fixed cube preserves
equal expected counts through repeated differencing, while second counting
moments control the dependence within the lists. Two coordinate searches
generate the lists and recover a shortest vector within these bounds.

The lattice kissing number gives a geometric interpretation of the factor
$3^n$. In the infinity norm, a lattice has at most $3^n-1$ shortest nonzero
vectors~\cite[Equation~(1.3)]{Henk02}, so $L\cap[-\lambda,\lambda]^n$
contains at most $3^n$ points, and $\Z^n$ attains this bound. The
terminal search makes the connection concrete through the disjoint
windows of \cref{fig:windows}, whose collected weight controls the
number of pair reports. On a general lattice the windows leave part of
the weight uncollected. When $L=\Z^n$ and $\ell=1$, every translate of
the half-open cube $[-1/2,1/2)^n$ contains exactly one lattice point,
so the windows tile the space and \eqref{eq:terminal-packing} becomes,
on an aligned grid, the equality
\[
 \sum_{w\in\Z^n}\prod_{i=1}^n c_\delta(w_i)=(4R_0-1)^n.
\]
This equals $3^n$ at $R_0=1$, and $3^n(1+o(1))$ for the algorithm's
choice $R_0=1+n^{-3}$.

Because $\Z^n$ has $3^n-1$ shortest vectors, \cref{thm:all-shortest}
achieves the optimal exponential running time in the worst case when all
shortest vectors must be listed explicitly. To solve SVP, however, any
one of those vectors suffices. On the extremal lattice $\Z^n$, the large
terminal report count comes with an abundance of shortest vectors,
leading to a natural question: can the abundance of shortest vectors be
exploited to find one without reporting all compatible pairs?

\appendix
\section{Basis preprocessing}\label{app:basis}
For completeness, we give the specialization of~\cite[Appendix~A.3]{GDD}
that proves \cref{lem:basis}, using the DBKZ theorem
of~\cite[Theorem~2.3]{ALNS20} and the exact Euclidean SVP algorithm
of~\cite{MV13}.

\begin{proof}[Proof of \cref{lem:basis}]
Suppose $n\ge4$ and put $k=\lfloor n/2\rfloor$.
Fix a prefix $F=(b_1,\ldots,b_i)$ of a basis $(F,D)$ of $\Lambda$.
Its orthogonal projection and remaining lattice are
\[
 \pi=I-F(F^{\mathsf T}F)^{-1}F^{\mathsf T},\qquad
 \Lambda_i=(\pi D)\Z^d=\pi\Lambda,\qquad d=n-i,
\]
with $\pi=I$ when $i=0$. Clear one common positive denominator in
$\pi D$ to obtain the integer input required by~\cite{ALNS20}.
This preserves length ratios and coefficients, so an output with
coefficient vector $z$ lifts to $Dz$.

If $d>k$, apply~\cite[Theorem~2.3]{ALNS20} with block dimension $k$,
exact oracle factor $\delta=1$, and slack $\varepsilon=1$. It returns
a nonzero projected vector $v$ with
\[
 \eunorm v\le
 2\gamma_k^{(d-1)/(2(k-1))}\det(\Lambda_i)^{1/d}
 \le2k^{(d-1)/(2(k-1))}\det(\Lambda_i)^{1/d},
\]
using polynomially many exact SVP calls of rank $k$.
Here $\gamma_k\le k$ is the Euclidean Hermite bound, and the
determinant is taken within the span of $\Lambda_i$.
If $d\le k$, use exact SVP on $\Lambda_i$.

Divide the output coefficient vector by the gcd of its coordinates
to make it primitive, then use the extended Euclidean algorithm to
obtain a unimodular completion. Lift this completion and size reduce
the selected column against the fixed prefix to obtain $b_{i+1}$, with
$\eunorm{b_{i+1}^*}\le\eunorm v$.
The projections of $n$ independent vectors of length at most
$\lambda_n^{(2)}(\Lambda)$ span the remaining space. Selecting $d$
independent projections and applying Hadamard's inequality gives
\[
 \det(\Lambda_i)^{1/d}\le\lambda_d^{(2)}(\Lambda_i)
 \le\lambda_n^{(2)}(\Lambda).
\]
Later operations keep the prefix fixed, so its Gram--Schmidt vectors
are preserved. Writing $n=2k+r$, $r\in\{0,1\}$, and using
$k^{1/(k-1)}\le2$ for $k\ge2$, we obtain
\[
 \max_i\eunorm{b_i^*}\le
 2k^{(n-1)/(2(k-1))}\lambda_n^{(2)}(\Lambda)
 =2k\,k^{(r+1)/(2(k-1))}\lambda_n^{(2)}(\Lambda)
 \le2n\lambda_n^{(2)}(\Lambda).
\]
For $n\le3$, exact SVP at every suffix gives the stronger factor one.
Thus $\Ared(n)=2(n+1)$ works in every dimension.

For the encoding bounds, begin with an LLL-reduced input basis~\cite{LLL82}.
After each selection, LLL reduce the projected suffix, lift its
integer column operations, and size reduce against earlier columns,
as in~\cite[Section~2]{MV13}. An LLL-reduced suffix has
Gram--Schmidt lengths at most $2^{n/2}\lambda_n^{(2)}(\Lambda)$.
Consequently every current basis column has length at most
$\sqrt{1+n/4}\max\{2n,2^{n/2}\}\lambda_n^{(2)}(\Lambda)$.
Its denominator divides the original common denominator, and
$\lambda_n^{(2)}(\Lambda)$ is at most the longest input column.
These bounds give polynomial encoding length at the start of each
round. Gram inverse and cofactor formulas give the same bound for
projections and integer coordinates. Each reduction, primitive
completion, and lift has polynomial intermediate size.

There are $n$ rounds and polynomially many exact SVP calls of rank
at most $k$. The algorithm of~\cite{MV13} uses $4^{k+o(k)}$ time
and $2^{k+o(k)}$ space per call, with polynomial input overhead.
The overall time and workspace are therefore bounded by
$2^{n+o(n)}\poly(\bits(B))$, including the fixed small dimensions.
\end{proof}

\section{Finite arithmetic and random choices}\label{app:implementation}
We record the finite representations used by the algorithm and the
coupling that makes every execution obey a deterministic resource bound.

\subsection{Constructing the superlattice and its quotient}
Lift $u\in\F_p^n$ to an integer vector. A Hermite basis of
\[
 K=p\Z^n+\Z u
\]
is obtained from $[pI\mid u]$ and gives a rational basis of $\Gamma$
after multiplication by $B/p$.

For a terminal entry $x=z+g$, let
\[
 a_x=pB^{-1}g\in\Z^n.
\]
Modulo $p$, this vector lies in the span of $u$. Choose an index $j$ with $u_j\ne0$ and represent the label by
\begin{equation}\label{eq:scalar-label}
 \lab(x)=(a_x)_j u_j^{-1}\pmod p.
\end{equation}
This scalar represents the class $g+L$, so equal labels are equivalent
to $x-y\in L$ and can be computed by exact rational and finite field arithmetic.

\subsection{Choosing the grid and sampling its quotient}
Choose the mesh in \eqref{eq:aligned-grid} from a common denominator
for $B/p$, $\ell/2$, and $R_j/6$, then subdivide until
$\delta/\ell\le n^{-10}$. Its encoding length is polynomial because
there are $O(\log n)$ radii of polynomial encoding length. The inclusion
$p^{-1}L\subseteq G$ makes this grid valid for every possible $\Gamma$
at the given scale and prime.

Let $H$ be an upper triangular integer Hermite basis of $\Gamma/\delta$. If its positive diagonal entries are $h_1,\ldots,h_n$, the integer box
\[
 \{0,\ldots,h_1-1\}\times\cdots\times\{0,\ldots,h_n-1\}
\]
is the standard set of representatives of $\Z^n/H\Z^n$.
Independent uniform coordinates in these intervals, multiplied by
$\delta$, therefore sample $G/\Gamma$ in polynomial time.

All node offsets can be reduced to these representatives after each subtraction without changing their cosets or the midpoint identities.

\subsection{Bounded randomness and bit complexity}
All random choices reduce to rejection sampling of uniform integers,
with acceptance probability at least $1/2$ per attempt.
The resource limits give a deterministic bound $R_{\mathrm{rand}}$
on the number of calls, with $\log R_{\mathrm{rand}}$
polynomial in the input length. Truncate each call after
\[
 s=\left\lceil\log_2R_{\mathrm{rand}}\right\rceil+n+20
\]
attempts and fail the trial if none succeeds. Coupling to
the exact choices gives a total disagreement probability at most
\[
 R_{\mathrm{rand}}2^{-s}\le2^{-n-20}.
\]
This includes quotient and field samples, prime candidates, and
rational retention coins.

Each record stores a vector in $Q_{R_t}\cap\delta\Z^n$, a quotient
label, and an identifier local to its node. Under the parameter and
resource limits, the records, the grid, the radii, and the table counts
all have polynomial encoding length, so the exact arithmetic, the
comparisons, and the membership tests $B^{-1}w\in\Z^n$ have polynomial
bit cost, as included in $p_{\mathrm{op}}$.

\section*{AI methodology}
\phantomsection
\label{sec:ai-methodology}
Building on Gaussian difference dynamics (GDD)~\cite{GDD}, the author
used GPT-6-astra to assist with extending the framework to the infinity
norm, including the algorithm design and mathematical analysis in
\crefrange{sec:cubes}{sec:algorithm}. The author then used GPT-6.1-sol
and GPT-6-astra to help draft the manuscript, and Claude Opus~5.5 and
Gemini~4 Carbon to help revise it. The Lean~4 formalization of
\cref{thm:main,thm:all-shortest} was developed with several AI models,
mainly Claude Opus~5.5, GPT-6.2-astra, GPT-6.2-sol, and Gemini~4 Carbon,
and all of its proofs are checked by the Lean~4 kernel. The author
checked the formal statements against the theorems, thoroughly checked
and revised the entire paper, and remains responsible for all
mathematical claims, proofs, and references.

\end{document}